\documentclass[10pt]{article}
\usepackage{amsmath}
\usepackage{graphicx}
\usepackage{color}
\usepackage[table,xcdraw]{xcolor}
\usepackage{doi}
\usepackage{orcidlink}
\usepackage{hyperref}
\hypersetup{colorlinks=true, linkcolor=blue, citecolor=blue, urlcolor=blue}
\usepackage{array}
\usepackage{longtable}
\newcommand{\ISCO}{ISCO}
\usepackage{diagbox}
\usepackage{booktabs}
\usepackage{colortbl}
\usepackage{amssymb}
\usepackage{subcaption}
\usepackage{geometry}
\renewcommand{\arraystretch}{1.8}
\RequirePackage[numbers,sort&compress]{natbib}
\newcommand{\BH}{BH}
\newcommand{\BHs}{BHs}
\newcommand{\GR}{GR}
\newcommand{\DM}{DM}
\newcommand{\CS}{CS}   
\newcommand{\CoS}{CoS} 
\newcommand{\GM}{GM}   
\newcommand{\QNM}{QNM}
\newcommand{\QNMs}{QNMs}
\newcommand{\WKB}{WKB}
\newcommand{\GBT}{GBT}
\newcommand{\PS}{PS}   
\newcommand{\EHT}{EHT}

\newcommand{\GF}{GF}   
\newcommand{\GFs}{GFs}
\newcommand{\EH}{EH}   

\begin{document}
\baselineskip=16pt

\begin{center}
{\Large Plummer Dark Matter Black Hole with Topological Defects: Shadow, \\[1ex] Greybody Factors,
Quasinormal Modes, and Thermodynamics}

\end{center}

\vspace{0.3cm}

\begin{center}
{\bf Ahmad Al-Badawi}\orcidlink{0000-0002-3127-3453}\\
Department of Physics, Al-Hussein Bin Talal University, 71111,
Ma'an, Jordan.\\
e-mail: ahmadbadawi@ahu.edu.jo\\
\vspace{0.2cm}
{\bf Faizuddin Ahmed}\orcidlink{0000-0003-2196-9622}\\
Department of Physics, The Assam Royal Global University,
Guwahati, 781035, Assam, India\\
e-mail: faizuddinahmed15@gmail.com\\
\vspace{0.2cm}
{\bf \.{I}zzet Sakall{\i}}\orcidlink{0000-0001-7827-9476}\\
Physics Department, Eastern Mediterranean University,
Famagusta 99628, North Cyprus via Mersin 10, Turkey\\
e-mail: izzet.sakalli@emu.edu.tr \,(Corresponding author)
\end{center}

\vspace{0.3cm}

\begin{abstract}
We construct a static, spherically symmetric black hole (BH) solution embedded in a cored Plummer dark matter (DM) halo and
a Letelier cloud of strings (CoS). Starting from the Plummer-Schwarzschild metric of Senjaya et al.~\cite{Senjaya2026}, we incorporate the string-cloud tension parameter $\alpha$ into the lapse function, obtaining $A(r) = h_{\rm Plummer}(r) - \alpha$. The resulting spacetime admits a single, non-degenerate event horizon (EH) for $\alpha < 1$ and a naked singularity for $\alpha \ge 1$. We determine the photon sphere (PS) and BH shadow radii, compute the weak deflection angle via the Gauss-Bonnet theorem (GBT), and analyze the innermost stable circular orbit (ISCO). Scalar perturbations are studied through the effective potential, greybody factor (GF) bounds obtained from the Boonserm-Visser method, the Hawking emission spectrum, and quasinormal mode (QNM) frequencies computed with the WKB approximation. The thermodynamic analysis covers the Hawking temperature, Bekenstein-Hawking entropy, heat capacity, and Gibbs free energy; the heat capacity is found to be strictly negative for all parameter values, confirming the absence of any Davies-type phase transition. A consistent hierarchy emerges across all six analyses: the CoS tension $\alpha$ governs the leading-order modifications to every observable, while the Plummer halo density $\rho_0$ provides a subdominant, additive correction.
\end{abstract}

\vspace{0.3cm}
\vspace{0.3cm}

\section{Introduction}\label{sec:1}

General relativity (\GR) has passed every precision test to date, from Solar System experiments to the direct detection of
gravitational waves by LIGO/Virgo~\cite{Abbott2016} and the first \BH\ shadow images obtained by the Event Horizon
Telescope (\EHT)~\cite{EHT2019}. Despite these triumphs, two major open questions persist: the nature of \DM\ on galactic
scales, and the role of topological defects-cosmic strings (\CS s), global monopoles (\GM s), and \CoS -generated in
early-universe symmetry-breaking phase transitions~\cite{Vilenkin1994}.

{\color{black}
A central question in the study of regular black holes or sourced by nonlinear electrodynamics or anisotropic fluids is whether such matter configurations can be realized in realistic astrophysical environments. In practice, black holes are never entirely isolated; they are embedded in large-scale structures often dominated by dark matter. Observational evidence from galaxies and clusters strongly supports the existence of dark matter halos \cite{NFW1997, Bertone2005}, although the precise microscopic nature of dark matter remains uncertain. To model these halos, phenomenological density profiles-such as those proposed by Hernquist \cite{Hernquist1990}, Einasto \cite{Einasto1965}, Dehnen \cite{Dehnen1993}, and the Dekel-Zhao model \cite{Dekel2017, Zhao1996}-are commonly employed in galactic dynamics and cosmological simulations. Notably, these profiles specify the density distribution while leaving the pressure unspecified, providing flexibility for effective fluid descriptions compatible with black hole solutions. In this work, we consider another DM halo profile known as Plummer profile~\cite{Plummer1911,Zeeuw1985} and investigate a black hole solution surrounded by a cloud of strings.} The density function of this Plummer profile~\cite{Plummer1911,Zeeuw1985} is givenby
\begin{equation}
\rho(r) = \frac{\rho_0}{\left[1+(r/r_0)^2\right]^2},
\label{plummer-density}
\end{equation}
stands out by combining a finite central density, a steep outer fall-off $\rho\sim r^{-4}$, and an analytically tractable
enclosed mass function. Senjaya et. al.~\cite{Senjaya2026} recently constructed the exact Schwarzschild-Plummer \BH\ solution and analyzed its photon dynamics, \QNMs\ in the eikonal limit, and thermodynamics~\cite{Senjaya2026}. Their metric function reads 
\begin{equation}
h_{\rm P}(r)
= \exp\!\left[-\frac{4\pi\rho_0 r_0^3}{r}
\tan^{-1}\!\left(\frac{r}{r_0}\right)\right]
- \frac{r_s}{r}\,,
\label{Plummer-profile}
\end{equation}
where $r_s = 2M$ is the Schwarzschild radius. This solution captures the gravitational effect of a cored \DM\ halo on the
central \BH\ geometry, but it does not account for topological defects that may coexist with the halo in realistic galactic
environments.

Topological defects provide an independent source of spacetime modification. A \CoS, first introduced by
Letelier~\cite{Letelier1979}, represents a distribution of one-dimensional objects threading spacetime. Its
energy-momentum tensor is parametrized by a single tension parameter $\alpha\in[0,1)$, which enters the metric function
as an additive constant $-\alpha$~\cite{Letelier1979,sec2is04}. A \GM, on the other hand, introduces a multiplicative
solid-deficit factor $(1-2\eta^{2})$~\cite{Barriola1989}, while a \CS\ generated by the Nambu-Goto action produces an
analogous conical-deficit modification~\cite{Vilenkin1981}. These defects arise naturally in grand-unified-theory phase
transitions and may thread the spacetime of supermassive \BHs\ at galactic centers, where \DM\ halos are also present.

The combined effect of a \DM\ halo and topological defects on \BH\ observables has attracted growing attention in recent
years~\cite{Rani2025,Gohain2024,AlBadawi2025,Ahmed2025,sec2is03,sec2is08}. Most existing studies, however, treat the two effects separately: either the \DM\ halo is studied in isolation, or a topological defect is superimposed on a vacuum \BH. The simultaneous incorporation of both a cored \DM\ profile and a string-cloud defect into a single \BH\ geometry, and the
investigation of its full phenomenological consequences, have not been carried out for the Plummer halo.

{\color{black}
The study of black hole thermodynamics has attracted considerable interest because it provides profound insights into the nature of black holes and their connection to the fundamental principles governing physical systems. Furthermore, it has significant implications for our understanding of quantum gravity. The foundational concepts of black hole thermodynamics were introduced in the seminal works \cite{Bardeen1973,Gibbons1977}, where the four laws of black hole mechanics were formulated, analogous to the laws of classical thermodynamics. Subsequent research revealed that black holes emit thermal radiation \cite{Hawking1975,Hawking1976} and possess an entropy proportional to the area of their event horizon \cite{Bekenstein1973,Bekenstein1974}. In black hole thermodynamics, various types of phase transitions have been studied. For instance, the Davies-type phase transition arises from a divergence in the heat capacity \cite{Davies1989}, while the Hawking–Page transition describes a phase change between thermal Anti-de Sitter (AdS) spacetime and black holes \cite{HawkingPage1983}. Other examples include extremal phase transitions \cite{PavonRubi1988,Pavon1991}, and Van der Waals–like behavior in the extended phase space, where the cosmological constant is treated as a thermodynamic pressure and the black hole mass is interpreted as enthalpy \cite{Kastor2009,KubiznakMann2012}.

Black holes return to equilibrium through damped oscillations characterized by their quasinormal modes (QNMs), making these complex frequencies valuable probes of strong-field gravity. The real part of a QNM corresponds to the oscillation frequency of the perturbation, whereas the imaginary part determines the rate of decay \cite{KonoplyaZhidenko2011, BertiCardoso2009}. Mathematically, QNMs appear as the eigenvalues of a Schrödinger-like wave equation subject to dissipative boundary conditions—ingoing at the event horizon and outgoing at spatial infinity \cite{Chandrasekhar1984, Leaver1985}. Because exact solutions are generally intractable for arbitrary potentials, QNMs are typically computed using semi-analytical or numerical methods, such as the WKB approximation, Padé summation, continued fractions, or time-domain integration \cite{Nollert1999, Matyjasek2019, CardosoPani2019}. The WKB method, first applied to Schwarzschild black holes in Refs. \cite{SchutzWill1985, Iyer1987, IyerWill1987}, provides a relatively simple yet effective semi-analytical approach for studying black hole perturbations. Later, this technique was extended to more general spacetimes, including rotating (Kerr) and charged (Reissner–Nordström) black holes \cite{KokkotasSchutz1988, Kokkotas1991}. Since then, QNMs have been investigated extensively across a wide range of black hole geometries and surrounding matter fields, including recent studies on regular and quantum-corrected black holes \cite{BonannoKonoplya2025, KonoplyaStashko2025}.
}

In this paper we fill this gap by constructing the Plummer-\CoS\ \BH\ solution-obtained by adding the Letelier
string-cloud parameter $\alpha$ to the Plummer-Schwarzschild metric of Ref.~\cite{Senjaya2026}-and studying its properties
across six interconnected domains: \EH\ structure, photon geodesics and shadow, weak gravitational lensing, timelike
geodesics and \ISCO, scalar perturbations (\GFs, \QNMs, and Hawking emission), and thermodynamics. Our analysis reveals
that the \CoS\ tension $\alpha$ and the Plummer halo density $\rho_{0}$ affect all observables in a correlated but
hierarchically ordered manner: $\alpha$ dominates the shifts in \EH\ radius, \PS, shadow size, \ISCO, \QNM\ frequencies, and thermodynamic quantities, while $\rho_{0}$ provides a secondary, additive correction. A notable structural feature of the solution is that it admits only a single, non-degenerate \EH\ for all $\alpha<1$, with no extremal or multi-horizon
configurations-a consequence of the absence of any repulsive barrier in the radial equation. We also show that the heat
capacity remains strictly negative for all parameter values, confirming that the Plummer-\CoS\ \BH\ is thermodynamically
unstable in the canonical ensemble, with no Davies-type phase transition.

The paper is organized as follows. In Sec.~\ref{sec:2} we construct the Plummer-\CoS\ metric, analyze its asymptotic
structure, and classify the \EH\ configurations across the full parameter space. Section~\ref{sec:3} studies null geodesics, the \PS, \BH\ shadow radius, effective radial force on photons, and weak deflection angle via the \GBT.
Section~\ref{sec:4} treats timelike geodesics and determines the \ISCO\ radius. In Sec.~\ref{sec:5} we derive the scalar
perturbation potential, compute \GF\ bounds using the Boonserm-Visser method, present the Hawking emission spectrum,
and obtain the \QNM\ frequencies via the \WKB\ approximation. Section~\ref{sec:6} covers the thermodynamic analysis: Hawking temperature, Bekenstein-Hawking entropy, heat capacity, and Gibbs free energy. We conclude in Sec.~\ref{sec:7}. Throughout we use natural units $G=c=\hbar=k_B=1$.

\section{Metric Construction and Horizon Structure}
\label{sec:2}

Astrophysical \BHs\ residing at galactic centers are surrounded by \DM\ halos whose gravitational imprint modifies the vacuum Schwarzschild geometry~\cite{sec2is01,sec2is02,sec2is03}. To model this environment analytically, one embeds the central \BH\ inside a matter distribution whose density profile $\rho(r)$ enters the metric through the enclosed mass function. Among the profiles studied in the literature-the cuspy Navarro-Frenk-White (NFW) form $\rho\propto r^{-1}(1+r/r_s)^{-2}$~\cite{NFW1997,Bertone2005}, and the cored pseudo-isothermal sphere-the Plummer profile~\cite{Plummer1911,Zeeuw1985} stands out by combining a finite central density with an analytically tractable enclosed mass. Simultaneously, topological defects produced during symmetry-breaking phase transitions in the early universe-\CS s, \GM s, and \CoS -leave permanent geometric imprints on the surrounding spacetime~\cite{Vilenkin1994,Letelier1979,Barriola1989}. In what follows we construct the Plummer-\CoS\ \BH\ line element, analyze its asymptotic structure, and classify its \EH\ configurations across the full parameter space.

\subsection{Plummer-Schwrazschild BH with Topological Defects: Plummer-\CoS\ \BH\ Spacetime}\label{subsec:plummer}

Recently, Senjaya et al.~\cite{Senjaya2026} obtained a static, spherically symmetric \BH\ solution immersed in a cored Plummer halo with density given by Eq.~(\ref{plummer-density}). The \DM\ mass enclosed within a radius $r$ reads
\begin{equation}
M_{\rm DM}(r) = 2\pi\rho_0 r_0^3 \left[ \tan^{-1}\!\left(\frac{r}{r_0}\right) - \frac{r\, r_0}{r^2 + r_0^2} \right],
\label{bb1}
\end{equation}
where $\rho_0$ is the central halo density and $r_0$ the core radius.

The tangential velocity relation $v_t^2 = M_{\rm DM}(r)/r$ then fixes the \DM-induced contribution to the metric function as~\cite{Senjaya2026,sec2is03}
\begin{equation}
f(r) = \exp\!\left[-\frac{4\pi\rho_0 r_0^3}{r}\,\tan^{-1}\!\left(\frac{r}{r_0}\right) \right],
\label{bb2}
\end{equation}
and the full Plummer-Schwarzschild lapse becomes $h(r) = f(r) - r_s/r$, with $r_s = 2M$. The corresponding line element takes the standard static, spherically symmetric form
\begin{equation}
ds^2 = -h(r)\, dt^2 + \frac{dr^2}{h(r)} + r^2 \left(d\theta^2 + \sin^2\theta \, d\phi^2\right).
\label{metric-1}
\end{equation}

Two properties of $f(r)$ deserve emphasis. First, as $r\to 0^+$ the argument of the exponential approaches $-4\pi\rho_0 r_0^2$ (since $\tan^{-1}(r/r_0)\sim r/r_0$ for small $r$), so that $f(0^+) = \exp(-4\pi\rho_0 r_0^2)$ remains finite; this is a direct consequence of the finite central density of the Plummer profile and contrasts sharply with the NFW case, where $\rho\to\infty$ as $r\to 0$~\cite{sec2is01}. Second, for $r\to\infty$ the exponential factor tends to unity, $f(\infty)=1$, recovering the Minkowski asymptotics of the seed Schwarzschild solution.

A \CoS\ in the Letelier model~\cite{Letelier1979} represents a collection of one-dimensional objects threading spacetime. The energy-momentum tensor of such a distribution is characterized by a single parameter $\alpha\in[0,1)$ that measures the string-cloud tension. Its net gravitational effect on a static, spherically symmetric geometry amounts to an additive downward shift of the lapse function~\cite{Letelier1979,sec2is04}:
\begin{equation}
h(r) \;\longrightarrow\; A(r) \equiv f(r) - \frac{r_s}{r} - \alpha
= \exp\!\left[-\frac{4\pi\rho_0 r_0^3}{r}\,\tan^{-1}\!\left(\frac{r}{r_0}\right) \right]-\frac{r_s}{r}-\alpha\,.
\label{function}
\end{equation}
The complete Plummer-\CoS\ \BH\ spacetime is then described by
\begin{equation}
ds^2 = -A(r)\, dt^2 + \frac{dr^2}{A(r)} + r^2 \left(d\theta^2 + \sin^2\theta \, d\phi^2\right),
\label{metric}
\end{equation}
with $A(r)$ given in Eq.~(\ref{function}). When $\alpha=0$ and $\rho_0=0$ the metric reduces to the Schwarzschild geometry; setting $\alpha=0$ alone recovers the Plummer-Schwarzschild solution of Ref.~\cite{Senjaya2026}; and keeping $\rho_0=0$ with $\alpha\neq 0$ reproduces the Letelier \BH~\cite{Letelier1979,sec2is04}.

The large-$r$ behaviour of the metric function is
\begin{equation}
A(r)\;\xrightarrow{r\to\infty}\; 1-\alpha\,,
\label{eq:asymptotic}
\end{equation}
because $f(\infty)=1$ and the $r_s/r$ term vanishes. Physical admissibility of the spacetime imposes a strict bound on the string-cloud parameter:
\begin{equation}
0\;\le\;\alpha\;<\;1\,.
\label{eq:alpha_bound}
\end{equation}
For $\alpha<1$ the asymptotic value $1-\alpha$ is positive, ensuring that $g_{tt}\to -(1-\alpha)<0$ and the spacetime signature remains Lorentzian at large distances. The metric is not asymptotically flat in the strict sense-since $A(\infty)\neq 1$-but can be brought to a manifestly flat form by rescaling the time coordinate $t\to t/\sqrt{1-\alpha}$~\cite{sec2is04,sec2is05}. This rescaling, commonly encountered in \CS\ and \CoS\ spacetimes, reflects the solid-angle deficit generated by the string distribution and does not affect the \EH\ locations, which are determined solely by the zeros of $A(r)$.

At the opposite extreme, $r\to 0^+$, the Schwarzschild pole $-r_s/r\to-\infty$ dominates over the bounded exponential and the constant $\alpha$, giving $A(0^+)\to -\infty$. Since $A(r)$ rises monotonically from $-\infty$ at the origin to the positive value $1-\alpha$ at spatial infinity, the intermediate-value theorem guarantees exactly one simple zero $r_h>0$ whenever $\alpha<1$. Furthermore, because this zero is simple-$A(r_h)=0$ and $A'(r_h)>0$-the horizon is always \emph{non-degenerate}. Unlike the Reissner-Nordstr\"{o}m or Kerr families, the Plummer-\CoS\ \BH\ admits neither an inner (Cauchy) horizon nor an extremal limit with degenerate horizons~\cite{sec2is06}. This single-horizon character follows from the fact that neither the cored \DM\ profile nor the string cloud introduces a repulsive (centrifugal or electromagnetic) barrier in the radial equation. 

When $\alpha\ge 1$, the asymptotic value $1-\alpha\le 0$ and the function $A(r)<0$ for all $r>0$, so no \EH\ forms. In this regime the metric describes a naked singularity, excluded by cosmic censorship arguments~\cite{sec2is07} and by the physical requirement $\alpha\ll 1$ for realistic string-cloud configurations~\cite{Vilenkin1994}.

\subsection{\EH\ equation and numerical results}
\label{subsec:horizon}

Setting $A(r_h)=0$ yields the implicit horizon equation
\begin{equation}
\exp\!\left[-\frac{4\pi\rho_0 r_0^3}{r_h}\,\tan^{-1}\!\left(\frac{r_h}{r_0}\right) \right]
= \frac{r_s}{r_h} + \alpha\,,
\label{eq:horizon_eq}
\end{equation}
which, owing to the transcendental nature of the left-hand side, cannot be solved in closed form. Analytical progress is possible in two limiting cases. When $\rho_0=0$ (no \DM), the exponential equals unity, and the horizon radius reduces to
\begin{equation}
r_h\big|_{\rho_0=0} = \frac{r_s}{1-\alpha}
= \frac{2M}{1-\alpha}\,,
\label{eq:rh_noDM}
\end{equation}
exhibiting the expected divergence as $\alpha\to 1^-$. For finite $\rho_0$ and small $\rho_0 r_0^3$, a first-order expansion of the exponential gives
\begin{equation}
r_h \;\approx\; \frac{2M}{1-\alpha}
\left[1 + \frac{4\pi\rho_0 r_0^3}{r_h^{(0)}}
\tan^{-1}\!\left(\frac{r_h^{(0)}}{r_0}\right)
+ \mathcal{O}\!\left(\rho_0^2\right)
\right],
\label{eq:rh_approx}
\end{equation}
where $r_h^{(0)} = 2M/(1-\alpha)$ is the $\rho_0=0$ result from Eq.~(\ref{eq:rh_noDM}). This expression shows that the \DM\ halo always \emph{increases} the horizon radius relative to the Schwarzschild-Letelier baseline, a physically expected result since the enclosed \DM\ mass adds to the gravitational pull.

For general parameter values, we solve Eq.~(\ref{eq:horizon_eq}) numerically using a sign-change root finder scanning $r\in[0.01,\,500\,M]$, followed by a refinement with the {\tt fsolve} routine in Maple~2024. The results are compiled in Table~\ref{tab:horizon-longtable}, which covers representative values of the halo parameters $\{\rho_0,\,r_0\}$ and the \CoS\ tension $\alpha$, including the naked singularity regime $\alpha\ge 1$.

\setlength{\tabcolsep}{10pt}
\renewcommand{\arraystretch}{1.5}
\begin{longtable}{|c|c|c|c|c|}
\hline
\rowcolor{orange!50}
\textbf{$\rho_0 M^2$} & \textbf{$r_0/M$} & \textbf{$\alpha$} & \textbf{$r_h/M$} & \textbf{Configuration} \\
\hline
\endfirsthead
\hline
\rowcolor{orange!50}
\textbf{$\rho_0 M^2$} & \textbf{$r_0/M$} & \textbf{$\alpha$} & \textbf{$r_h/M$} & \textbf{Configuration} \\
\hline
\endhead
0.0 & 0.2 & 0.00 & $[2.0000]$ & Single horizon BH \\
\hline
0.5 & 0.2 & 0.00 & $[2.0728]$ & Single horizon BH \\
\hline
1.0 & 0.2 & 0.00 & $[2.1435]$ & Single horizon BH \\
\hline
0.0 & 0.2 & 0.10 & $[2.2222]$ & Single horizon BH \\
\hline
0.0 & 0.2 & 0.30 & $[2.8571]$ & Single horizon BH \\
\hline
0.5 & 0.2 & 0.10 & $[2.3038]$ & Single horizon BH \\
\hline
0.5 & 0.2 & 0.30 & $[2.9637]$ & Single horizon BH \\
\hline
1.0 & 0.2 & 0.30 & $[3.0681]$ & Single horizon BH \\
\hline
0.5 & 0.5 & 0.10 & $[3.2718]$ & Single horizon BH \\
\hline
0.0 & 0.2 & 0.50 & $[4.0000]$ & Single horizon BH \\
\hline
0.5 & 0.2 & 0.70 & $[6.9236]$ & Single horizon BH \\
\hline
0.0 & 0.2 & 0.90 & $[20.000]$ & Single horizon BH \\
\hline
0.5 & 0.2 & 0.95 & $[41.573]$ & Single horizon BH \\
\hline
0.0 & 0.2 & 1.00 & $[\,]$ & Naked singularity \\
\hline
0.5 & 0.2 & 1.20 & $[\,]$ & Naked singularity \\
\hline
\caption{\EH\ radius $r_h/M$ and spacetime configuration for various Plummer \DM\ halo parameters $\{\rho_0,\,r_0\}$ and \CoS\ tension $\alpha$, with $M=1$. The notation $[\,]$ denotes the absence of a horizon (naked singularity). Numerical values are obtained by solving $A(r_h)=0$ with a sign-change root finder in Maple.}
\label{tab:horizon-longtable}
\end{longtable}

Several features emerge from Table~\ref{tab:horizon-longtable}. First, the Schwarzschild result $r_h=2M$ is recovered in the first row ($\rho_0=\alpha=0$). Second, the horizon radius grows with both $\rho_0$ and $\alpha$: the Plummer halo adds enclosed \DM\ mass, while the \CoS\ reduces the asymptotic lapse, both pushing the zero of $A(r)$ to larger radii. For instance, adding $\rho_0 M^2=0.5$ at $\alpha=0$ shifts the horizon from $2M$ to $2.0728\,M$-a modest $3.6\%$ increase driven entirely by the enclosed \DM\ mass. In contrast, setting $\alpha=0.3$ at $\rho_0=0$ produces a much larger displacement to $r_h\approx 2.857\,M$, indicating that the \CoS\ tension is the dominant horizon-shifting mechanism in the astrophysically accessible regime. Third, as $\alpha$ approaches unity, the horizon radius increases rapidly: $r_h\approx 20\,M$ at $\alpha=0.9$ and $r_h\approx 41.6\,M$ at $\alpha=0.95$ (with $\rho_0 M^2=0.5$), consistent with the divergence predicted by Eq.~(\ref{eq:rh_noDM}). Fourth, the rightmost column confirms the absence of any horizon for $\alpha\ge 1$, in agreement with the analytic argument of Sec.~\ref{sec:2}. The transition from a single-horizon \BH\ to a naked singularity is therefore a sharp, first-order phenomenon controlled by the string-cloud parameter. The entry at $r_0/M=0.5$ (ninth row) demonstrates that a larger core radius also produces a larger horizon ($r_h=3.2718\,M$ versus $2.3038\,M$ at $r_0/M=0.2$), since the \DM\ mass enclosed within a given radius grows with $r_0$.

In addition to Table~\ref{tab:horizon-longtable}, which surveys a broad parameter space-including different core radii $r_0/M\in\{0.2,\,0.5\}$, extreme string-cloud tensions up to $\alpha=0.95$, and naked singularity configurations---we present in Table~\ref{tab:1} a finer two-parameter scan of $r_h$ over $\alpha\in[0,0.3]$ and $\rho_0 M^2\in[0,0.3]$ at fixed $r_0/M=0.2$, which captures the parameter region most relevant for astrophysically motivated models where $\alpha\ll 1$ and $\rho_0 r_0^3\ll M$~\cite{sec2is08}. While Table~\ref{tab:horizon-longtable} classifies the spacetime configurations qualitatively, Table~\ref{tab:1} resolves the quantitative interplay between $\alpha$ and $\rho_0$ on a fine grid, making the relative weight of each parameter immediately visible.

\begin{table}[ht!]
\centering
\begin{tabular}{|c|c|c|c|c|c|c|c|}
\hline
\rowcolor{orange!50}
\diagbox[innerwidth=2.8cm, height=1.1cm, linecolor=black]{\textbf{\hspace{12mm}$\alpha$}}{ \textbf{$\rho_0 M^2$}} & \textbf{0.00} & \textbf{0.05} & \textbf{0.10} & \textbf{0.15} & \textbf{0.20} & \textbf{0.25} & \textbf{0.30} \\
\hline
0.00 & 2.00000 & 2.00738 & 2.01474 & 2.02208 & 2.02939 & 2.03668 & 2.04395 \\
\hline
0.05 & 2.10526 & 2.11306 & 2.12084 & 2.12859 & 2.13632 & 2.14402 & 2.15171 \\
\hline
0.10 & 2.22222 & 2.23048 & 2.23872 & 2.24693 & 2.25512 & 2.26329 & 2.27143 \\
\hline
0.15 & 2.35294 & 2.36172 & 2.37047 & 2.37920 & 2.38790 & 2.39658 & 2.40524 \\
\hline
0.20 & 2.50000 & 2.50936 & 2.51869 & 2.52800 & 2.53728 & 2.54654 & 2.55578 \\
\hline
0.25 & 2.66667 & 2.67668 & 2.68667 & 2.69664 & 2.70658 & 2.71650 & 2.72639 \\
\hline
0.30 & 2.85714 & 2.86791 & 2.87865 & 2.88937 & 2.90006 & 2.91073 & 2.92138 \\
\hline
\end{tabular}
\caption{\EH\ radius $r_h/M$ as a function of $\alpha$ and $\rho_0$, with fixed $r_s/M=2$ and $r_0/M=0.2$. For each column $\rho_0$ increases the horizon by a few percent at most, whereas each row (increasing $\alpha$) produces a substantially larger shift.}
\label{tab:1}
\end{table}

\subsection{Metric function behavior and graphical analysis}\label{subsec:metric-plot}

\begin{figure}[ht!]
    \centering
    \includegraphics[width=0.95\linewidth]{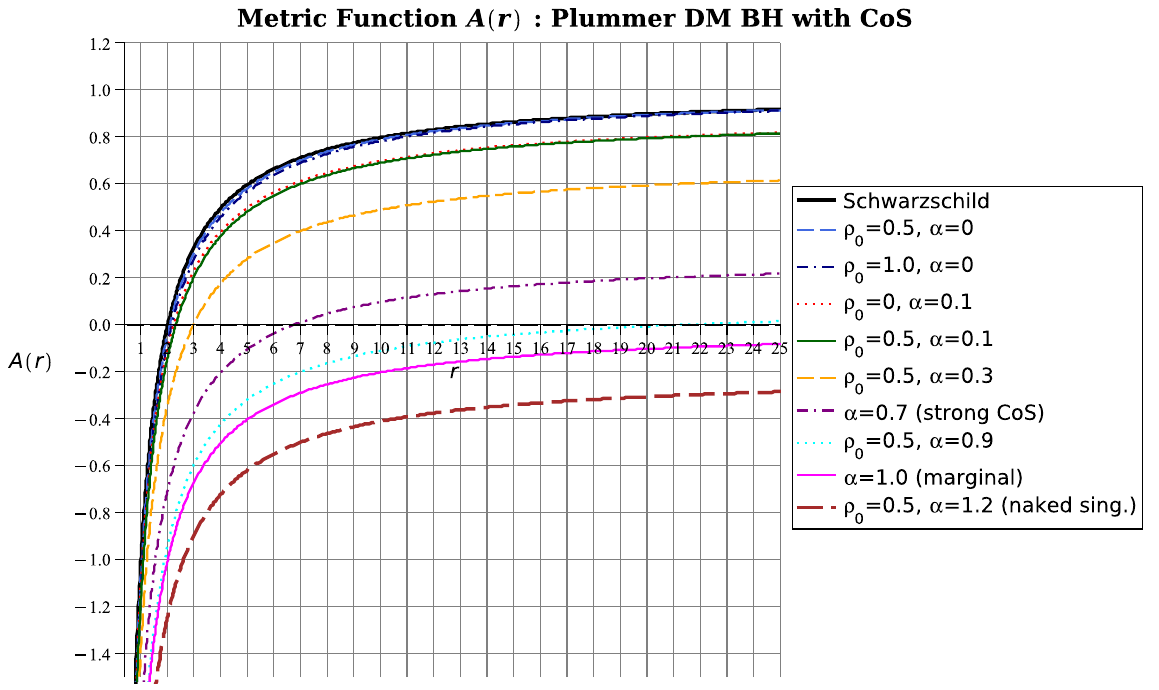}
    \caption{Metric function $A(r)$ for various Plummer \DM\ and \CoS\ parameter combinations. The black dashed horizontal line marks $A=0$. Curves that cross this line possess a single \EH; those remaining entirely below it (e.g., $\alpha=1.0$ and $\alpha=1.2$) correspond to naked singularities. Parameters: $M=1$, $r_0/M=0.2$.}
    \label{fig:metric-all}
\end{figure}

Figure~\ref{fig:metric-all} displays the metric function $A(r)$ for a selection of parameter combinations that illustrate the full range of spacetime configurations. The solid black curve represents the Schwarzschild baseline ($\rho_0=\alpha=0$), which crosses zero at $r=2M$. Blue curves show the effect of the Plummer \DM\ halo alone: the exponential suppression of $f(r)$ at intermediate radii lowers the curve and pushes the zero crossing to slightly larger $r$. Red and purple curves isolate the \CoS\ effect: increasing $\alpha$ shifts the entire curve downward by a constant amount, reducing the asymptotic value from $1$ to $1-\alpha$ and moving the horizon to progressively larger radii. The green and orange curves show the combined \DM\ + \CoS\ case, where both effects add constructively. The near-marginal configuration $\alpha=0.95$ (cyan) has its horizon pushed beyond $r\sim 40\,M$, while the brown dashed curve ($\alpha=1.2$) lies entirely below the zero line, confirming the naked singularity regime. The marginal case $\alpha=1.0$ (magenta) approaches $A(\infty)=0$ from below, so $A(r)<0$ everywhere and no horizon exists.

\begin{figure}[ht!]
    \centering
    \includegraphics[width=0.45\linewidth]{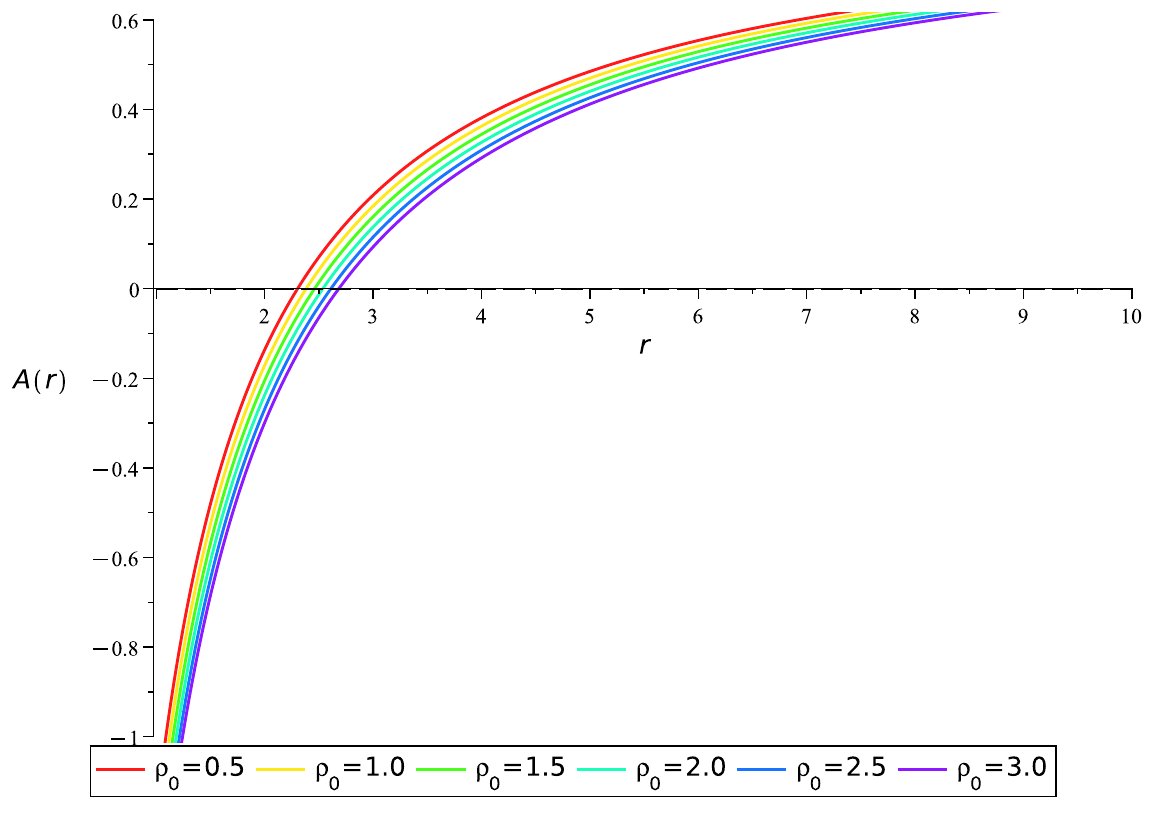}\qquad
    \includegraphics[width=0.45\linewidth]{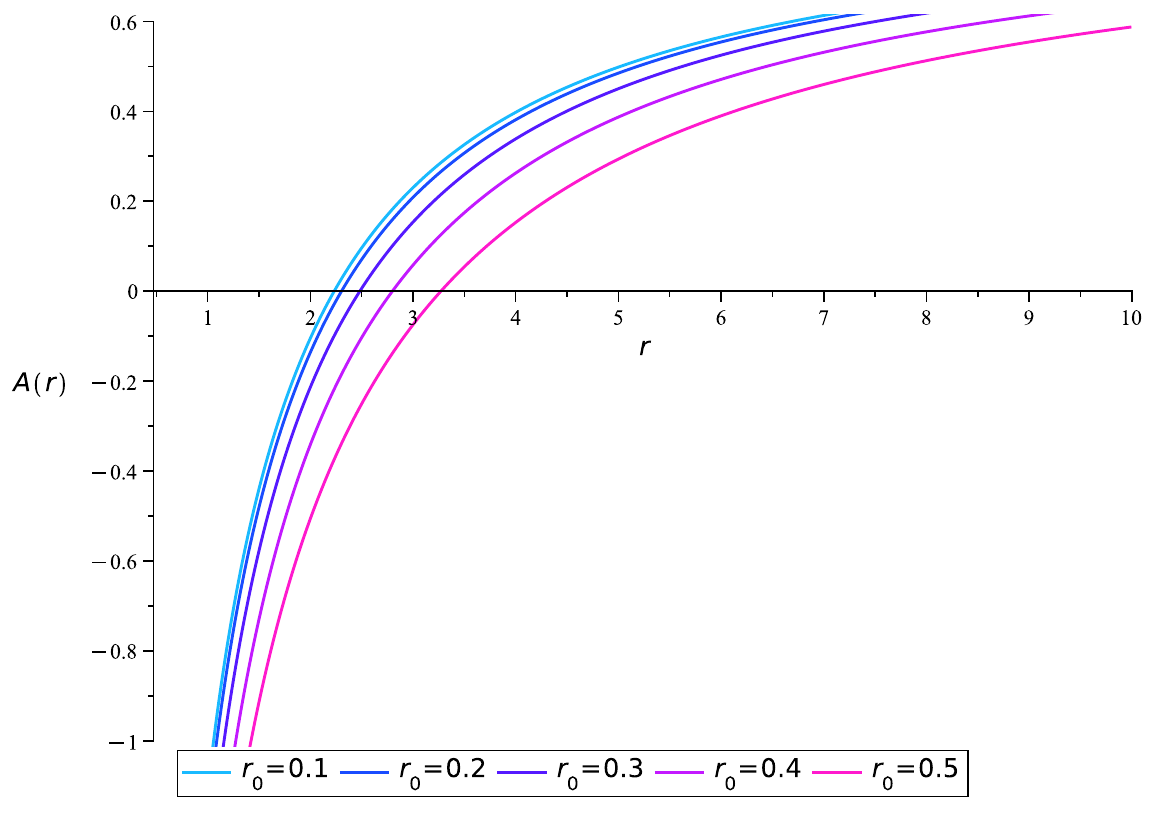}\\
    (i) $r_0/M=0.2,\;\alpha=0.1$ \hspace{5cm} (ii) $\rho_0=0.5/M^2,\;\alpha=0.1$\\[3ex]
    \includegraphics[width=0.65\linewidth]{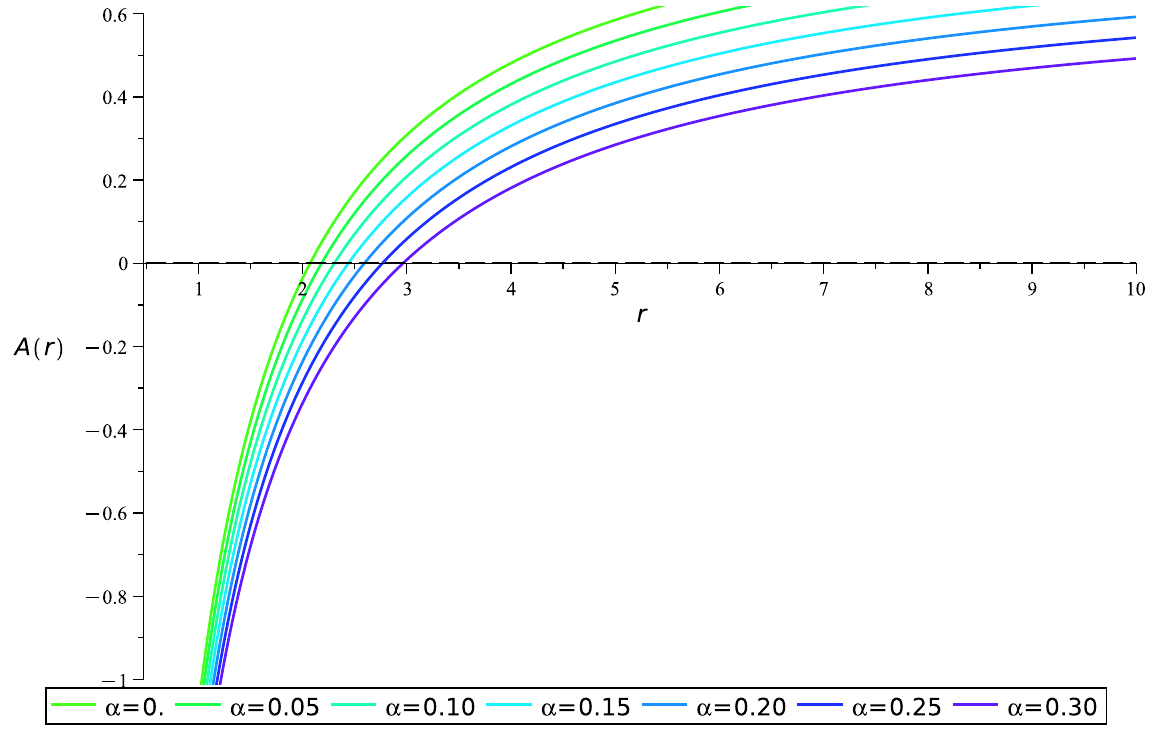}\\
    (iii) $r_0/M=0.2,\;\rho_0=0.5/M^2$
    \caption{Metric function $A(r)$ as a function of the dimensionless radial distance for various halo parameters $\{r_0,\,\rho_0\}$ and \CoS\ tension $\alpha$. Panel~(i): varying $\rho_0$ at fixed $r_0/M=0.2$ and $\alpha=0.1$. Panel~(ii): varying $r_0$ at fixed $\rho_0 M^2=0.5$ and $\alpha=0.1$. Panel~(iii): varying $\alpha$ at fixed $r_0/M=0.2$ and $\rho_0 M^2=0.5$. The zero crossing of each curve marks the corresponding \EH\ radius.}
    \label{fig:metric-function}
\end{figure}

The three-panel parameter study in Fig.~\ref{fig:metric-function} complements the above by displaying separately the dependence on $\rho_0$ (panel~i), $r_0$ (panel~ii), and $\alpha$ (panel~iii), each with the remaining parameters held fixed. In panel~(i), increasing the central halo density $\rho_0$ at fixed $r_0/M=0.2$ and $\alpha=0.1$ progressively lowers $A(r)$ in the region $r\sim 2$--$5\,M$ and shifts the horizon outward, while the far-field value $A(\infty)=1-\alpha=0.9$ remains unchanged. Panel~(ii) shows that enlarging the core radius $r_0$ has a qualitatively similar effect: a wider halo core encloses more \DM\ mass at moderate radii, deepening the exponential suppression and again pushing the horizon to larger $r$. Panel~(iii) varies $\alpha$ at fixed $\rho_0$ and $r_0$; here the entire curve shifts rigidly downward, confirming the additive nature of the \CoS\ contribution. In all three panels the peak of $A(r)$ decreases as the varied parameter grows, confirming that both the \DM\ halo and the \CoS\ weaken the effective gravitational barrier experienced by test fields and particles propagating in this geometry. These trends \cite{Sakalli2025Letelier,Ahmed2026PhotonLV,Ahmed2025QC,Sakalli2025YM,Sakalli2025Thermo,Sadeghi2024WGC} will carry over to the \PS, shadow, and \QNM\ analyses of the subsequent sections.

We close this section by noting that the \BH\ mass can be expressed in terms of the horizon radius by inverting $A(r_h)=0$:
\begin{equation}
M = \frac{r_h}{2}\left[\exp\!\left\{-\frac{4\pi\rho_0 r_0^3}{r_h}\,\tan^{-1}\!\left(\frac{r_h}{r_0}\right)\right\} - \alpha\right].
\label{eq:mass_rh}
\end{equation}
This relation will serve as the starting point for the thermodynamic analysis in Sec.~\ref{sec:6}, where the \EH\ radius $r_h$ plays the role of the thermodynamic coordinate.


\section{Null Geodesics: Shadow and Weak Lensing}
\label{sec:3}

The causal structure of a spacetime can be probed by studying the geodesic motion of test particles and photons~\cite{Chandrasekhar1984,Wald1984}. This analysis gives access to the \PS, the \BH\ shadow boundary, and the weak-field deflection angle-three quantities that encode the combined gravitational imprint of the Plummer \DM\ halo and the \CoS\ on photon propagation. The geodesic equation $\left(ds/d\lambda\right)^2=\varepsilon$ (with $\varepsilon=-1$ for massive particles and $\varepsilon=0$ for photons) applied to the metric~(\ref{metric}) yields
\begin{equation}
-A(r)\,\dot{t}^{2}+\frac{\dot{r}^{2}}{A(r)}+r^{2}\,\dot{\theta}^{2}+r^{2}\sin^{2}\!\theta\;\dot{\phi}^{2}=\varepsilon\,,
\label{cc1}
\end{equation}
where the dot denotes differentiation with respect to the affine parameter $\lambda$. The two Killing symmetries of the metric-stationarity and axial symmetry-lead to the conserved specific energy $\mathrm{E}$ and specific angular momentum $\mathrm{L}$:
\begin{equation}
\dot{t}=\frac{\mathrm{E}}{A(r)}\,,
\qquad
\dot{\phi}=\frac{\mathrm{L}}{r^{2}}\,.
\label{cc2}
\end{equation}
Substituting Eq.~(\ref{cc2}) into~(\ref{cc1}) and restricting to equatorial photon orbits ($\varepsilon=0$, $\theta=\pi/2$) gives the radial equation of motion
\begin{equation}
\dot{r}^{2}+\frac{\mathrm{L}^{2}}{r^{2}}\,A(r)=\mathrm{E}^{2}\,,
\label{cc3}
\end{equation}
which has the form of one-dimensional motion with energy $\mathrm{E}^{2}$ in an effective potential $V_{\rm eff}(r)=\mathrm{L}^{2}\,A(r)/r^{2}$.

\subsection{Effective potential and \PS}
\label{subsec:effective-potential}

The null effective potential reads
\begin{equation}
V_{\rm eff}(r)
=\frac{\mathrm{L}^{2}}{r^{2}}\,A(r)
=\frac{\mathrm{L}^{2}}{r^{2}}
\left[\exp\!\left\{-\frac{4\pi\rho_{0}r_{0}^{3}}{r}
\tan^{-1}\!\left(\frac{r}{r_{0}}\right)\right\}
-\frac{r_{s}}{r}-\alpha\right].
\label{cc4}
\end{equation}
The shape of $V_{\rm eff}$ determines whether a photon is captured by the \EH, scattered to infinity, or temporarily trapped on a circular orbit. The peak of $V_{\rm eff}$ defines the unstable circular photon orbit, i.e.\ the \PS.

\begin{figure}[ht!]
    \centering
    \includegraphics[width=0.45\linewidth]{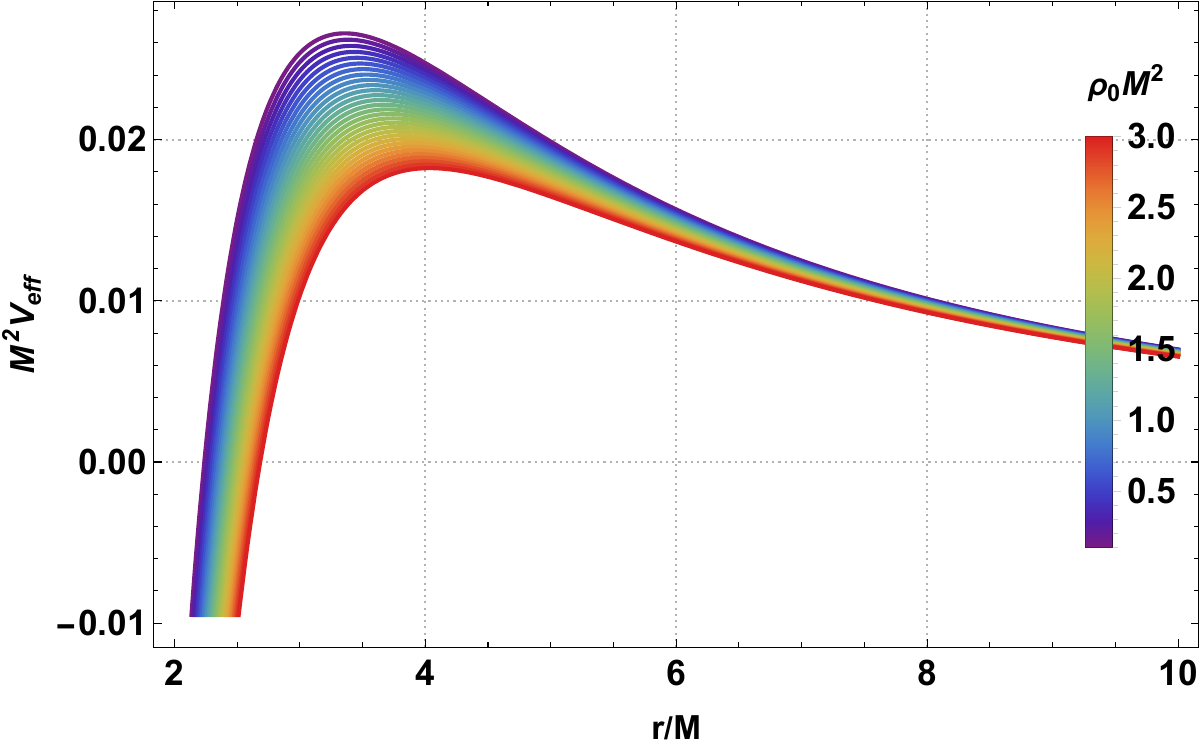}\qquad
    \includegraphics[width=0.45\linewidth]{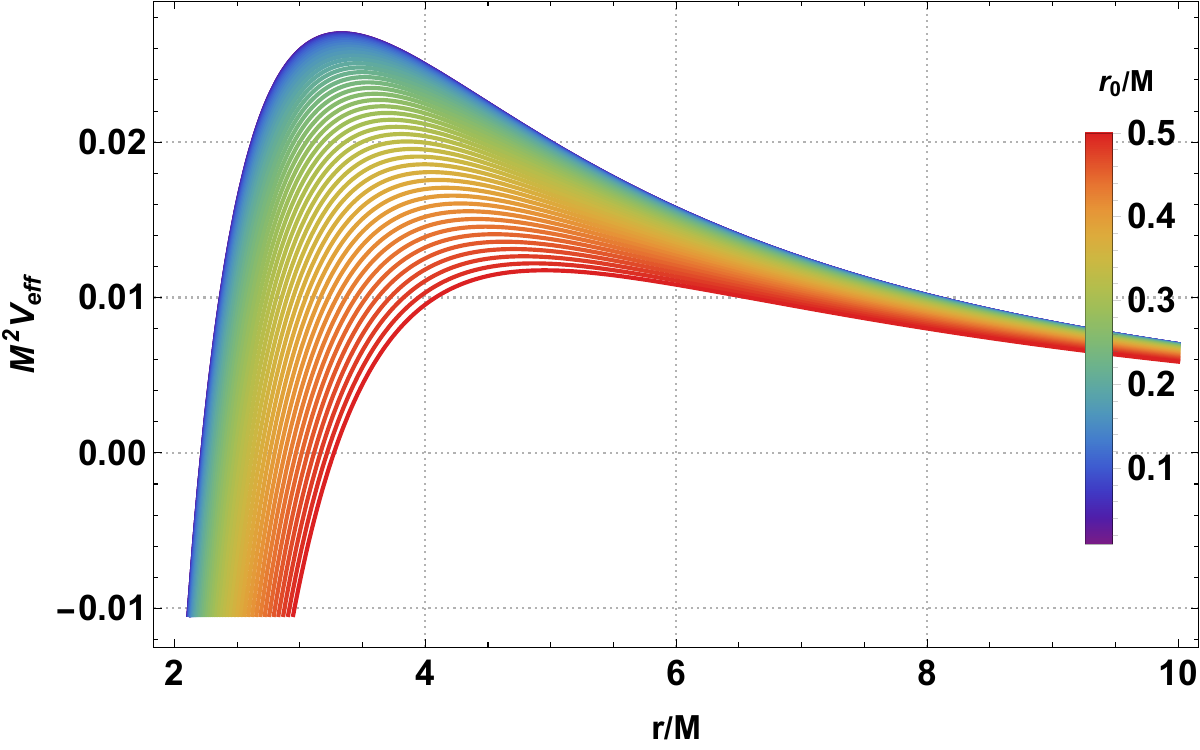}\\
    (i) $r_0/M=0.2,\;\alpha=0.1$ \hspace{4cm} (ii) $\rho_0=0.5/M^2,\;\alpha=0.1$\\[3ex]
    \includegraphics[width=0.45\linewidth]{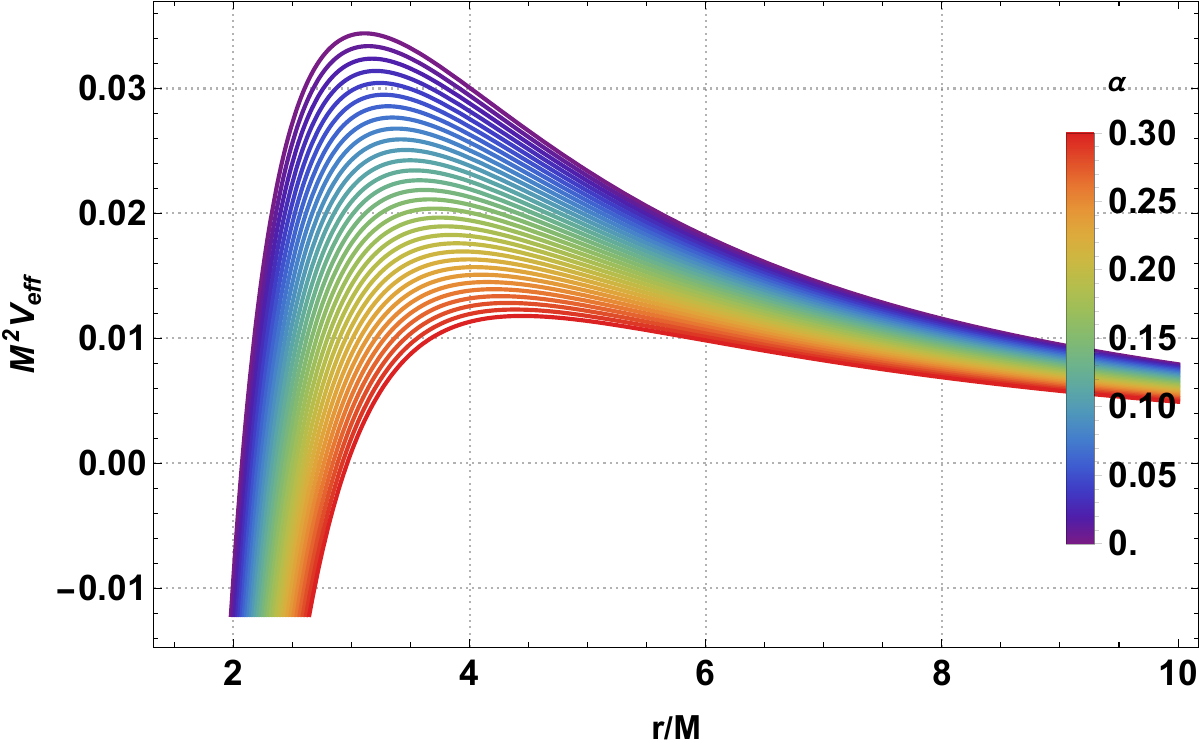}\\
    (iii) $r_0/M=0.2,\;\rho_0=0.5/M^2$
    \caption{Null effective potential $V_{\rm eff}(r)$ as a function of the dimensionless radial coordinate for various Plummer halo parameters $\{r_0,\,\rho_0\}$ and \CoS\ tension $\alpha$. In all panels the peak decreases and moves outward as the varied parameter grows, signaling a weakening of the photon trapping barrier. Here $\mathrm{L}/M=1$.}
    \label{fig:potential-null}
\end{figure}

In Fig.~\ref{fig:potential-null} we plot $V_{\rm eff}(r)$ for different values of the Plummer halo parameters $\{\rho_{0},\,r_{0}\}$ and the \CoS\ tension $\alpha$. Panel~(i) varies $\rho_{0}$ at fixed $r_{0}/M=0.2$ and $\alpha=0.1$: higher central densities reduce the peak height and shift it outward, meaning that the potential barrier weakens and the \PS\ moves to larger radii. Panel~(ii) varies $r_{0}$ at fixed $\rho_{0}M^{2}=0.5$ and $\alpha=0.1$, producing a qualitatively similar trend-a wider \DM\ core encloses more mass at intermediate radii and lowers the barrier. Panel~(iii) varies $\alpha$ at fixed $\rho_{0}$ and $r_{0}$; here the entire potential shifts downward by an amount proportional to $\alpha$, reflecting the additive nature of the \CoS\ contribution in~(\ref{function}). In every case the barrier height decreases monotonically, indicating that photons are more weakly bound as either the \DM\ or \CoS\ content increases.

Circular photon orbits of radius $r_{\rm ph}$ satisfy $V_{\rm eff}(r_{\rm ph})=\mathrm{E}^{2}$ and $V_{\rm eff}'(r_{\rm ph})=0$ simultaneously~\cite{Chandrasekhar1984}, which reduces to the condition
\begin{equation}
2\,A(r_{\rm ph})-r_{\rm ph}\,A'(r_{\rm ph})=0\,.
\label{cc6}
\end{equation}
Substituting $A(r)$ from~(\ref{function}), this becomes
\begin{equation}
\exp\!\left[-\frac{4\pi\rho_{0}r_{0}^{3}}{r}
\tan^{-1}\!\left(\frac{r}{r_{0}}\right)\right]
\left[1+2\pi\rho_{0}r_{0}^{3}
\left\{\frac{r_{0}}{r^{2}+r_{0}^{2}}
-\frac{1}{r}\tan^{-1}\!\left(\frac{r}{r_{0}}\right)\right\}\right]
-\alpha-\frac{3r_{s}}{2r}=0\,,
\label{cc8}
\end{equation}
which is transcendental in $r$ and must be solved numerically. In the Schwarzschild limit ($\rho_{0}=\alpha=0$) the equation reduces to $1-3r_{s}/(2r)=0$, giving the well-known result $r_{\rm ph}=3M$. For finite $\rho_{0}$ or $\alpha$ the \PS\ radius increases beyond this value, as confirmed by the numerical data compiled in Table~\ref{tab:photon-sphere-radius}.

\begin{table}[ht!]
\centering
\begin{tabular}{|c|c|c|c|c|c|c|c|}
\hline
\rowcolor{orange!50}
\diagbox[innerwidth=2.2cm, height=1.2cm, linecolor=black]{\textbf{$\rho_0 M^2$}}{\textbf{$\alpha$}} & \textbf{0.0} & \textbf{0.05} & \textbf{0.1} & \textbf{0.15} & \textbf{0.2} & \textbf{0.25} & \textbf{0.3} \\
\hline
0.0 & 3.0000 & 3.1579 & 3.3333 & 3.5294 & 3.7500 & 4.0000 & 4.2857 \\
\hline
0.1 & 3.0462 & 3.2073 & 3.3864 & 3.5867 & 3.8120 & 4.0674 & 4.3595 \\
\hline
0.2 & 3.0922 & 3.2566 & 3.4394 & 3.6438 & 3.8738 & 4.1348 & 4.4331 \\
\hline
0.3 & 3.1380 & 3.3057 & 3.4921 & 3.7007 & 3.9356 & 4.2020 & 4.5067 \\
\hline
0.4 & 3.1836 & 3.3546 & 3.5448 & 3.7575 & 3.9972 & 4.2691 & 4.5803 \\
\hline
0.5 & 3.2291 & 3.4033 & 3.5972 & 3.8142 & 4.0587 & 4.3361 & 4.6537 \\
\hline
\end{tabular}
\caption{\PS\ radius $r_{\rm ph}/M$ for different values of $\rho_0 M^2$ and $\alpha$, with $r_0/M=0.2$. The Schwarzschild value $r_{\rm ph}=3M$ is recovered at $\rho_0=\alpha=0$. Both parameters increase $r_{\rm ph}$ monotonically, with $\alpha$ producing the larger effect.}
\label{tab:photon-sphere-radius}
\end{table}

Table~\ref{tab:photon-sphere-radius} reveals two trends. Along each row, increasing $\alpha$ from $0$ to $0.3$ enlarges the \PS\ by roughly $40\%$ (e.g., from $3M$ to $4.286\,M$ at $\rho_{0}=0$), while along each column, increasing $\rho_{0}M^{2}$ from $0$ to $0.5$ produces only a $\sim 7\%$ growth (e.g., from $3M$ to $3.229\,M$ at $\alpha=0$). The \CoS\ tension therefore dominates the shift of the \PS, a pattern consistent with the horizon analysis of Sec.~\ref{subsec:horizon}.

\subsection{Shadow radius}
\label{subsec:shadow}

Black hole shadows represent the apparent dark region observed against a bright background, resulting from the extreme bending of light by the black hole's gravitational field. Photons approaching the black hole either fall into the event horizon or escape to infinity, and the unstable photon orbits, known as the photon sphere, determine the shadow's boundary \cite{Chandrasekhar1984,Wald1984}. Studying black hole shadows enables constraints on black hole parameters and tests for the general relativity, as well as investigations into the influence of surrounding matter such as DM halos and topological defects. Several works on black hole shadow has been explored in the literature (see, \cite{Uniyal2026,Lin2025,Sultana2026}). Here, we show how DM halo as well as string clouds modify the shadow size in comparison to the standard black hole.

The critical impact parameter for photon capture follows from the conditions~(\ref{cc6}) and reads~\cite{Perlick2022}
\begin{equation}
b_{c}=\frac{r_{\rm ph}}{\sqrt{A(r_{\rm ph})}}\,.
\label{dd1}
\end{equation}
For a static observer located at $r\to\infty$, the angular radius of the \BH\ shadow is~\cite{Perlick2022}
\begin{equation}
R_{\rm sh}=b_{c}\,\sqrt{A(r\to\infty)}
=b_{c}\,\sqrt{1-\alpha}\,,
\label{dd2}
\end{equation}
where we used $A(\infty)=1-\alpha$ from Eq.~(\ref{eq:asymptotic}). Substituting Eq.~(\ref{dd1}) gives the explicit expression
\begin{equation}
R_{\rm sh}
=\frac{\sqrt{1-\alpha}\;r_{\rm ph}}
{\sqrt{\exp\!\left[-\dfrac{4\pi\rho_{0}r_{0}^{3}}{r_{\rm ph}}
\tan^{-1}\!\left(\dfrac{r_{\rm ph}}{r_{0}}\right)\right]
-\dfrac{r_{s}}{r_{\rm ph}}-\alpha}}\,.
\label{dd4}
\end{equation}
Setting $\alpha=0$ recovers the Schwarzschild-Plummer shadow of Ref.~\cite{Senjaya2026}; further setting $\rho_{0}=0$ yields the Schwarzschild result $R_{\rm sh}=3\sqrt{3}\,M\approx 5.196\,M$.

\begin{table}[ht!]
\centering
\begin{tabular}{|c|c|c|c|c|c|c|c|}
\hline
\rowcolor{orange!50}
\diagbox[innerwidth=2.2cm, height=1.2cm, linecolor=black]{\textbf{$\rho_0 M^2$}}{\vspace{8mm} \textbf{$\alpha$}} & \textbf{0.0} & \textbf{0.05} & \textbf{0.1} & \textbf{0.15} & \textbf{0.2} & \textbf{0.25} & \textbf{0.3} \\
\hline
0.0 & 5.1962 & 5.4696 & 5.7735 & 6.1131 & 6.4952 & 6.9282 & 7.4231 \\
\hline
0.1 & 5.3736 & 5.6593 & 5.9769 & 6.3321 & 6.7319 & 7.1853 & 7.7038 \\
\hline
0.2 & 5.5504 & 5.8484 & 6.1799 & 6.5508 & 6.9685 & 7.4425 & 7.9850 \\
\hline
0.3 & 5.7266 & 6.0371 & 6.3826 & 6.7694 & 7.2052 & 7.7001 & 8.2667 \\
\hline
0.4 & 5.9024 & 6.2254 & 6.5851 & 6.9879 & 7.4420 & 7.9579 & 8.5489 \\
\hline
0.5 & 6.0776 & 6.4134 & 6.7873 & 7.2062 & 7.6789 & 8.2160 & 8.8317 \\
\hline
\end{tabular}
\caption{Shadow radius $R_{\rm sh}/M$ for different values of $\rho_0 M^2$ and $\alpha$, with $r_0/M=0.2$. The Schwarzschild value $R_{\rm sh}\approx 5.196\,M$ appears at $\rho_0=\alpha=0$. Both parameters enlarge the shadow monotonically.}
\label{tab:shadow-radius}
\end{table}

\begin{figure}[ht!]
    \centering
    \includegraphics[width=0.45\linewidth]{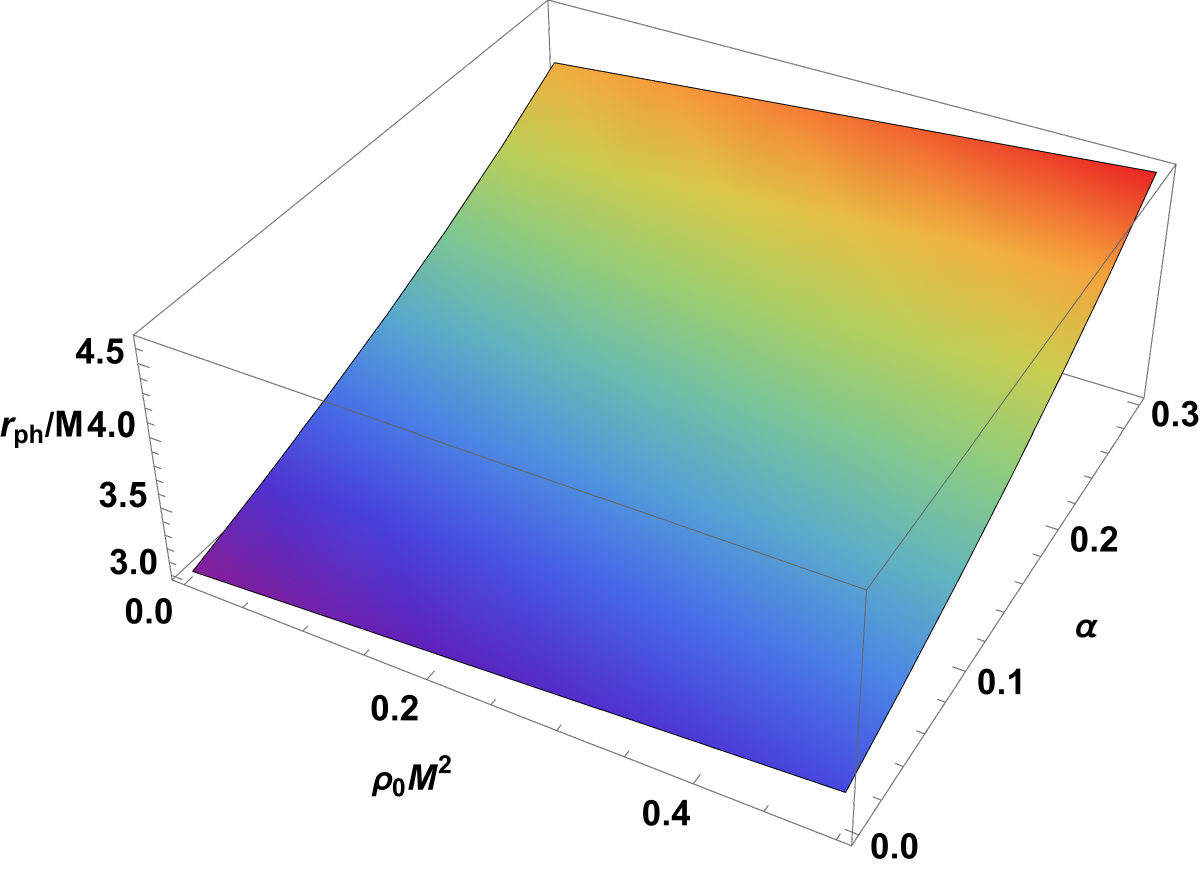}\qquad
    \includegraphics[width=0.45\linewidth]{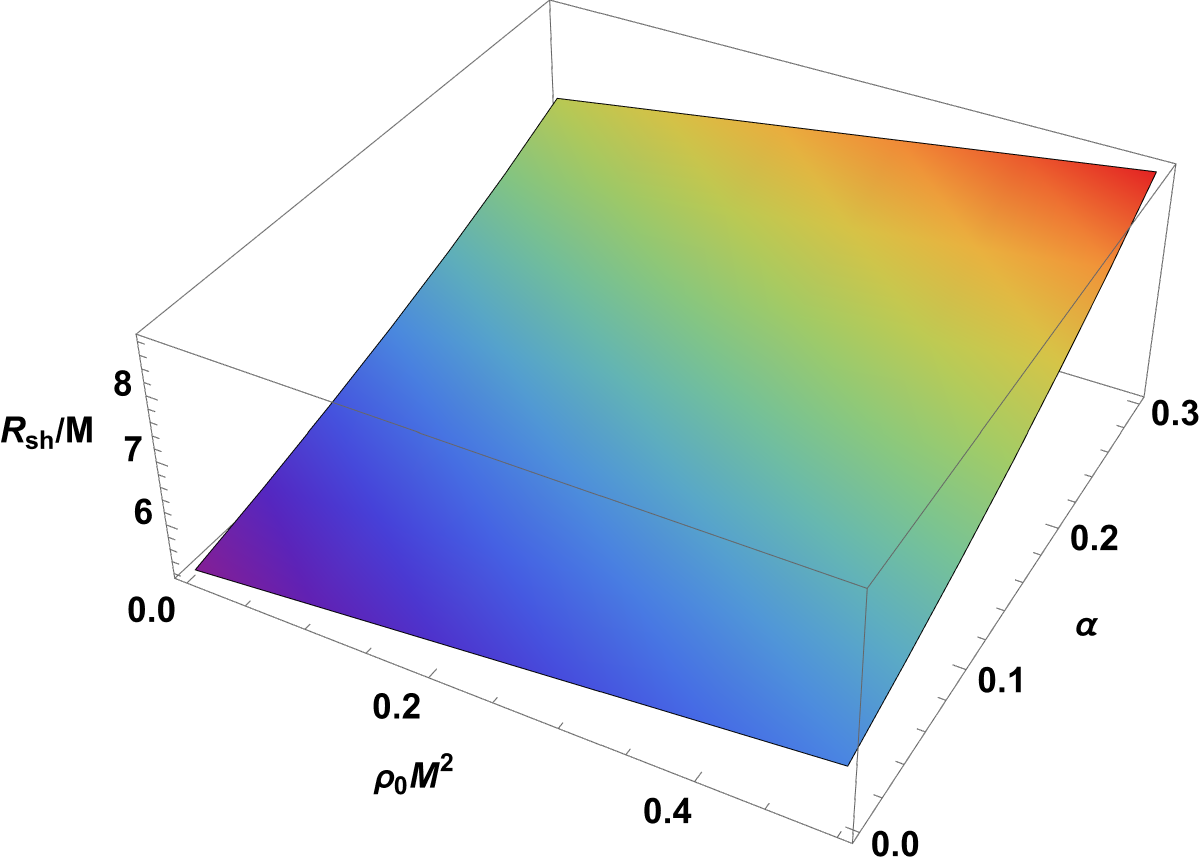}
    \caption{\PS\ radius $r_{\rm ph}/M$ (left) and shadow radius $R_{\rm sh}/M$ (right) as functions of $\{\rho_0,\,\alpha\}$ at fixed core radius $r_0/M=0.2$. Both surfaces increase monotonically with $\rho_0$ and $\alpha$, with the steeper gradient along the $\alpha$-axis reflecting the dominant role of the \CoS\ tension.}
    \label{fig:photon-shadow}
\end{figure}

Table~\ref{tab:shadow-radius} lists the shadow radius for the same parameter grid used in Table~\ref{tab:photon-sphere-radius}. In line with the \PS\ data, $R_{\rm sh}$ grows with both $\rho_{0}$ and $\alpha$, and the \CoS\ tension again accounts for the dominant contribution: at $\rho_{0}M^{2}=0.5$ and $\alpha=0.3$, the shadow reaches $R_{\rm sh}\approx 8.83\,M$, roughly $70\%$ larger than the Schwarzschild baseline. The three-dimensional parameter dependence is visualized in Fig.~\ref{fig:photon-shadow}, where both $r_{\rm ph}$ and $R_{\rm sh}$ are plotted as surfaces over the $(\rho_{0},\alpha)$ plane at fixed $r_{0}/M=0.2$. The surfaces rise steeply as $\alpha$ increases, confirming the \CoS\ as the dominant shadow-enlarging mechanism. These shadow predictions can in principle be confronted with the \EHT\ measurements of M87$^{*}$ and Sgr~A$^{*}$~\cite{EHT2019,EHT2022}.

\subsection{Effective radial force on photons}
\label{subsec:force}

Further information about the photon dynamics is encoded in the effective radial force, defined as the negative gradient of $V_{\rm eff}$~\cite{Chandrasekhar1984}:
\begin{equation}
\mathcal{F}=-\frac{1}{2}\,\frac{\partial V_{\rm eff}}{\partial r}\,.
\label{dd5}
\end{equation}
Substituting the potential~(\ref{cc4}) yields
\begin{equation}
\mathcal{F}=\frac{\mathrm{L}^{2}}{r^{3}}
\left[\exp\!\left(-\frac{4\pi\rho_{0}r_{0}^{3}}{r}
\tan^{-1}\!\left(\frac{r}{r_{0}}\right)\right)
\left\{1+2\pi\rho_{0}r_{0}^{3}
\left(\frac{r_{0}}{r^{2}+r_{0}^{2}}
-\frac{1}{r}\tan^{-1}\!\left(\frac{r}{r_{0}}\right)
\right)\right\}
-\alpha-\frac{3r_{s}}{2r}\right].
\label{eq:force}
\end{equation}
The zeros of $\mathcal{F}$ coincide with the extrema of $V_{\rm eff}$: the outermost zero corresponds to the \PS, where the inward gravitational pull on the photon exactly balances the centrifugal repulsion. For $r<r_{\rm ph}$ the force is directed inward ($\mathcal{F}<0$), while for $r>r_{\rm ph}$ a narrow region of outward force exists before $\mathcal{F}$ decays to zero at large $r$.

\begin{figure}[ht!]
    \centering
    \includegraphics[width=0.45\linewidth]{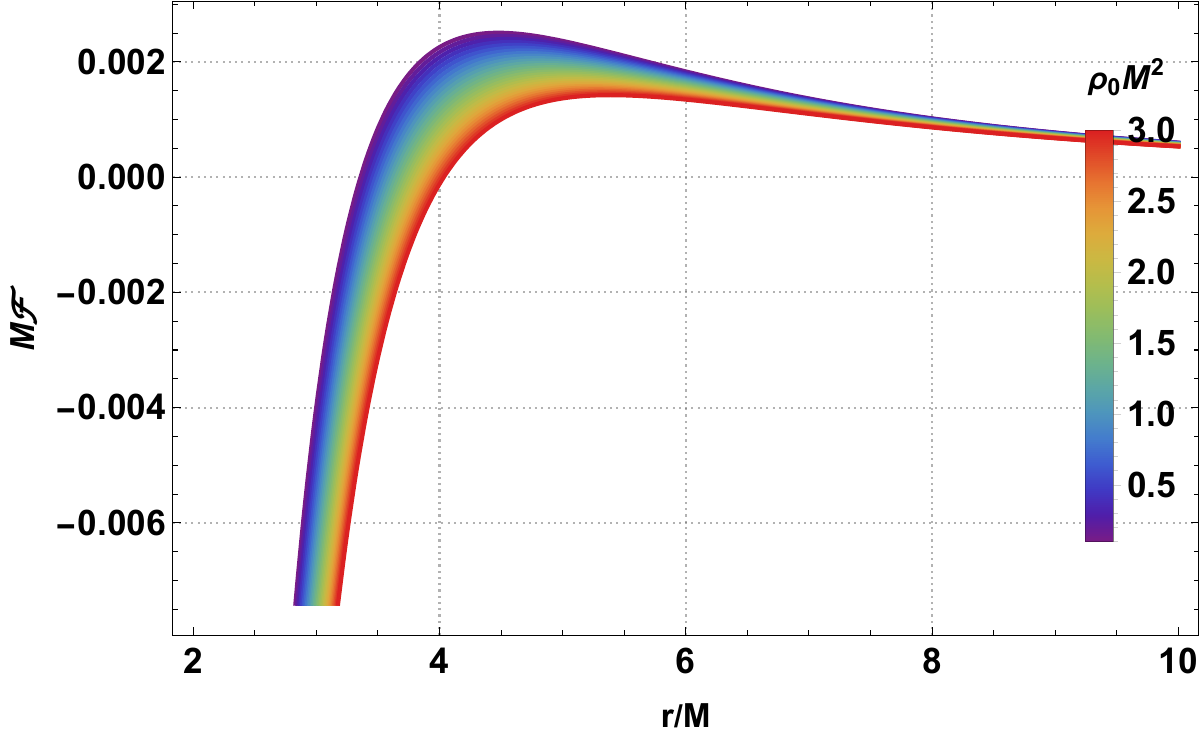}\qquad
    \includegraphics[width=0.45\linewidth]{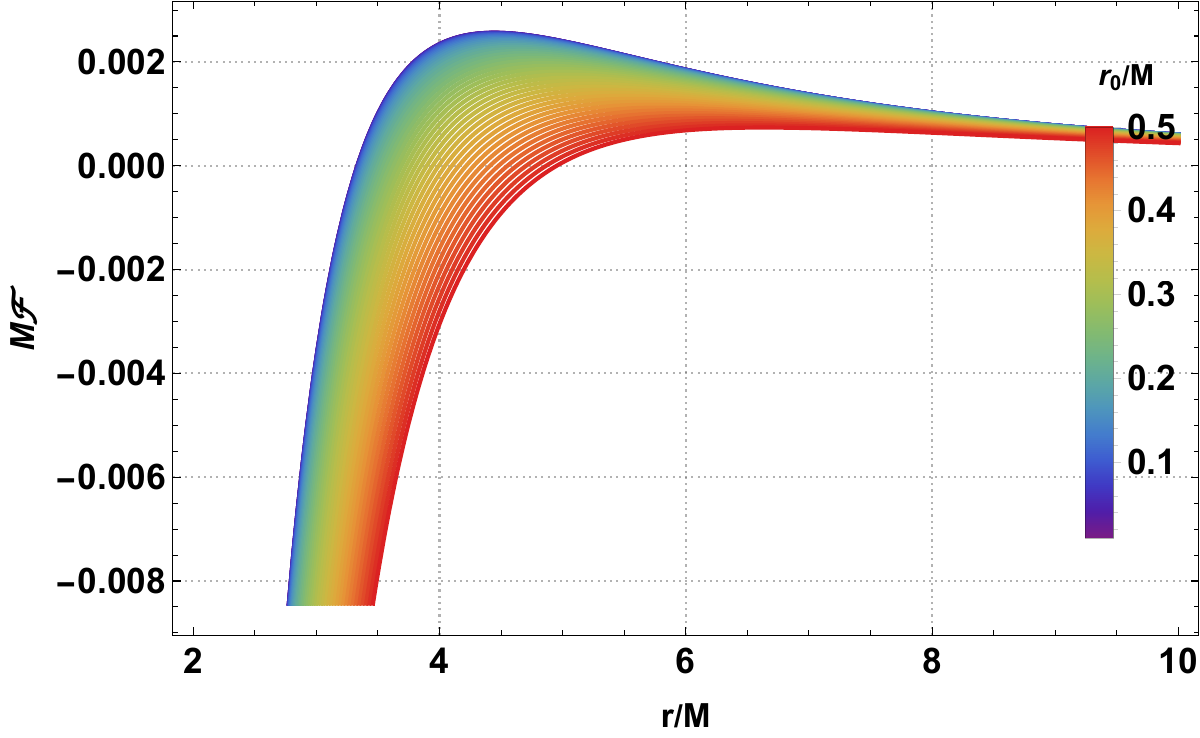}\\
    (i) $r_0/M=0.2,\;\alpha=0.1$ \hspace{4cm} (ii) $\rho_0=0.5/M^2,\;\alpha=0.1$\\[3ex]
    \includegraphics[width=0.45\linewidth]{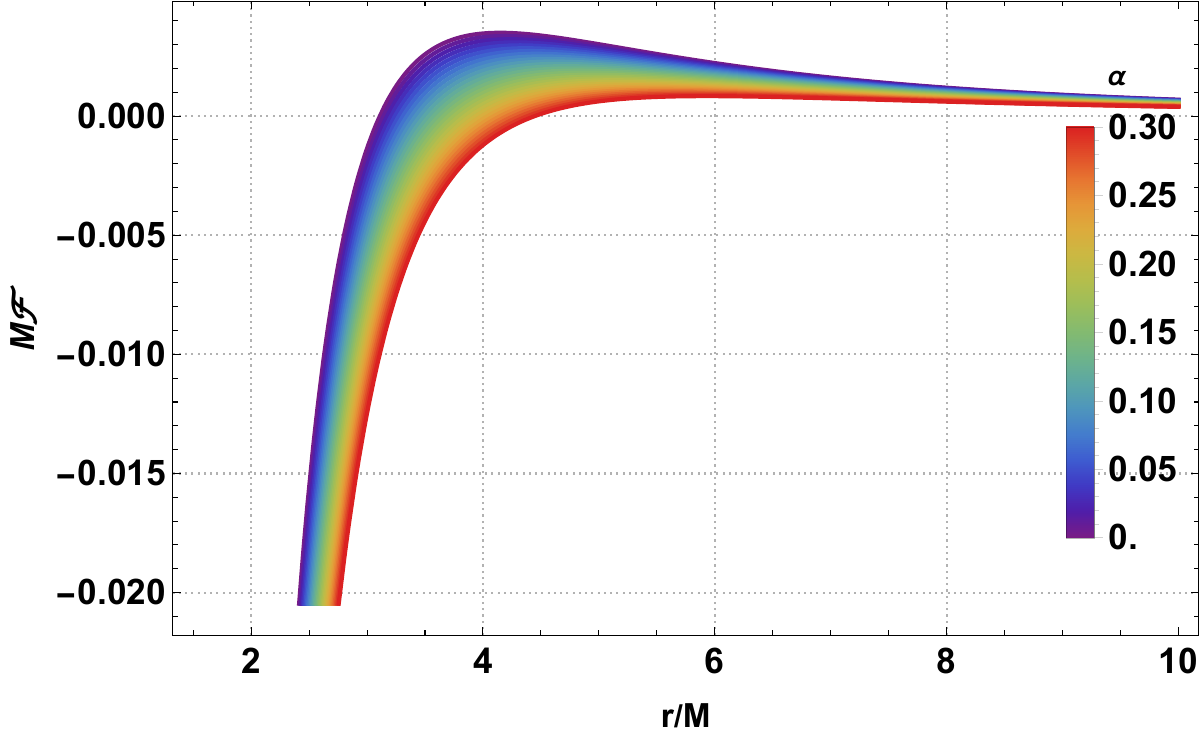}\\
    (iii) $r_0/M=0.2,\;\rho_0=0.5/M^2$
    \caption{Effective radial force $\mathcal{F}$ experienced by photons as a function of the dimensionless radial coordinate for various Plummer halo parameters $\{r_0,\,\rho_0\}$ and \CoS\ tension $\alpha$. The zero crossing of each curve marks the \PS\ radius $r_{\rm ph}$. Here $\mathrm{L}/M=1$.}
    \label{fig:force}
\end{figure}

Figure~\ref{fig:force} displays $\mathcal{F}(r)$ for the same parameter variations used in the potential plots. In all three panels the magnitude of the radial force decreases as $\rho_{0}$, $r_{0}$, or $\alpha$ increases, consistent with the weakening of the effective potential barrier observed in Fig.~\ref{fig:potential-null}. Physically, a stronger \DM\ halo or a denser \CoS\ softens the gravitational grip on photons orbiting near the \PS, making the unstable circular orbit easier to escape.

\subsection{Weak deflection angle via the \GBT}
\label{subsec:GBT}

We compute the weak-field gravitational deflection angle using the \GBT\ formulation of Gibbons and Werner~\cite{GibbonsWerner2008}. The optical metric associated with the line element~(\ref{metric}) is obtained by setting $ds^{2}=0$ and reads
\begin{equation}
d\sigma^{2}=\frac{dr^{2}}{A^{2}(r)}
+\frac{r^{2}\,d\varphi^{2}}{A(r)}\,.
\label{eq:optical_metric}
\end{equation}
The Gaussian curvature $\mathcal{K}$ of this two-dimensional Riemannian manifold is~\cite{GibbonsWerner2008,sec3is01}
\begin{equation}
\mathcal{K}=\frac{1}{2}\left[
\frac{1}{2}\left(\frac{dA}{dr}\right)^{\!2}
-A(r)\,\frac{d^{2}A}{dr^{2}}\right].
\label{eq:GaussCurv}
\end{equation}
The \GBT\ applied to a spatial domain $\mathcal{D}_{R}$ bounded by the photon ray and a circular arc of radius $R\to\infty$ gives the deflection angle~\cite{GibbonsWerner2008}
\begin{equation}
\hat{\alpha}_{\rm def}
=-\iint_{\mathcal{D}_{R}}\mathcal{K}\,dS\,,
\label{eq:GBT_integral}
\end{equation}
where $dS$ is the area element of the optical metric. For a non-asymptotically flat spacetime with $A(\infty)=1-\alpha\neq 1$, the integration requires careful treatment of the boundary terms~\cite{Ahmed2026PhotonLV,Ahmed2025QC}. Following the procedure of Refs.~\cite{Sakalli2025KZ,Gashti2026WGC}, we expand $A(r)$ to leading order in $r_{s}/r$ and $\rho_{0}r_{0}^{3}/r$ and evaluate the integral~(\ref{eq:GBT_integral}). In the large-$r$ regime, $\tan^{-1}(r/r_{0})\to\pi/2$, so the \DM-induced term in $A(r)$ behaves as $-2\pi^{2}\rho_{0}r_{0}^{3}/r$, contributing an effective mass-like correction. The resulting weak deflection angle for a photon with impact parameter $b$ is
\begin{equation}
\hat{\alpha}_{\rm def} \approx\frac{4M}{(1-\alpha)\,b} +\frac{4\pi^{2}\rho_{0}r_{0}^{3}}{(1-\alpha)\,b}
+\frac{3\pi^{3}\rho_{0}r_{0}^{3}\,r_{s}}{2(1-\alpha)\,b^{2}}+\mathcal{O}\!\left(\frac{r_{s}^{2}}{b^{2}},\,\rho_{0}^{2}\right).
\label{eq:alpha_CoS}
\end{equation}
The first term is the standard Einstein deflection modified by the \CoS\ through the factor $(1-\alpha)^{-1}$; this enhancement arises because the conical deficit produced by the string cloud effectively magnifies the apparent gravitational mass~\cite{sec3is02,Letelier1979}. The second term is a purely \DM-induced deflection proportional to the enclosed Plummer mass at large distances, where $\tan^{-1}(r/r_{0})\to\pi/2$. The third term captures the leading-order cross-coupling between the \DM\ halo and the Schwarzschild mass. In the limit $\rho_{0}\to 0$, Eq.~(\ref{eq:alpha_CoS}) reduces to $4M/[(1-\alpha)b]$, which reproduces the Schwarzschild--Letelier deflection~\cite{sec3is04,sec3is02}. Setting $\alpha=0$ instead recovers the Plummer--Schwarzschild deflection of Ref.~\cite{Senjaya2026}.

Equation~(\ref{eq:alpha_CoS}) shows that the \DM\ halo and the \CoS\ both increase the deflection angle relative to the Schwarzschild baseline, but through distinct mechanisms: the Plummer profile adds a genuine mass term, while the \CoS\ rescales the entire deflection by $(1-\alpha)^{-1}$ without introducing new matter content. This structural difference means that, in principle, independent measurements of the deflection angle and the shadow radius could disentangle the two contributions, since $R_{\rm sh}$ depends on $\sqrt{1-\alpha}$ [Eq.~(\ref{dd2})] while $\hat{\alpha}_{\rm def}$ depends on $(1-\alpha)^{-1}$.

\section{Timelike Geodesics and \ISCO\ Analysis}
\label{sec:4}

The motion of massive test particles around a \BH\ determines the structure of accretion disks, the efficiency of energy extraction, and the frequencies of quasi-periodic oscillations observed in X-ray binaries~\cite{sec4is01,sec4is02}. A central quantity in this context is the \ISCO, which marks the innermost radius at which a stable circular orbit exists. Below the \ISCO, no bound circular motion is possible and infalling matter plunges directly toward the \EH. For a Schwarzschild \BH\ the \ISCO\ lies at $r_{\rm ISCO}=6M$; modifications of the geometry-whether by \DM\ or topological defects-shift this radius and thereby alter the observable signatures of the accretion flow.

The Lagrangian for a test particle of mass $m$ in the spacetime~(\ref{metric}) is
\begin{equation}
\mathbb{L}=\frac{m}{2}\,g_{\mu\nu}\,
\frac{dx^{\mu}}{d\lambda}\,\frac{dx^{\nu}}{d\lambda}\,.
\label{timelike-1}
\end{equation}
The Killing symmetries yield two conserved quantities-the specific energy $\mathcal{E}$ and the specific angular momentum $\mathcal{L}$-together with the equations of motion
\begin{align}
\frac{dt}{d\lambda}&=\frac{\mathcal{E}}{A(r)}\,,
\label{timelike-2}\\[4pt]
\frac{d\phi}{d\lambda}&=\frac{\mathcal{L}}{r^{2}\sin^{2}\theta}\,,
\label{timelike-3}\\[4pt]
\frac{d\theta}{d\lambda}&=\frac{p_{\theta}}{m\,r^{2}}\,,
\label{timelike-4}\\[4pt]
\left(\frac{dr}{d\lambda}\right)^{2}
+\frac{p_{\theta}^{2}}{m\,r^{2}}\,A(r)
+U_{\rm eff}(r,\theta)&=\mathcal{E}^{2}\,,
\label{timelike-5}
\end{align}
where $p_{\theta}$ is the $\theta$-component of the four-momentum. The effective potential governing the radial motion reads
\begin{equation}
U_{\rm eff}(r,\theta)
=\left(1+\frac{\mathcal{L}^{2}}{r^{2}\sin^{2}\theta}\right)A(r)\,.
\label{potetial-timelike}
\end{equation}

\subsection{Effective potential for massive particles}
\label{subsec:potential-massive}

The structure of $U_{\rm eff}$ is controlled by the Plummer halo parameters $\{\rho_{0},\,r_{0}\}$, the \CoS\ tension $\alpha$, the \BH\ mass $M$, and the particle angular momentum $\mathcal{L}$. At large $r$ the potential approaches $A(\infty)=1-\alpha$, so a test particle at rest at infinity has $\mathcal{E}^{2}=1-\alpha<1$ rather than unity; this is a direct consequence of the conical deficit introduced by the string cloud.

\begin{figure}[ht!]
    \centering
    \includegraphics[width=0.45\linewidth]{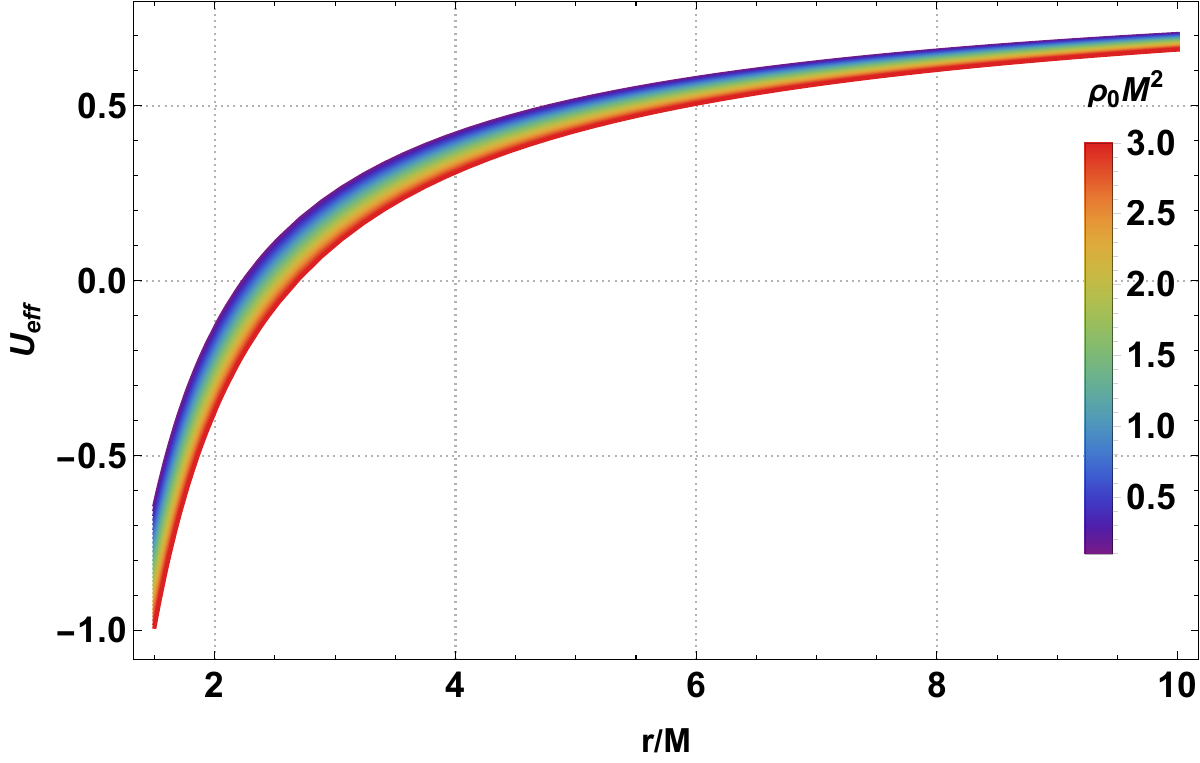}\qquad
    \includegraphics[width=0.45\linewidth]{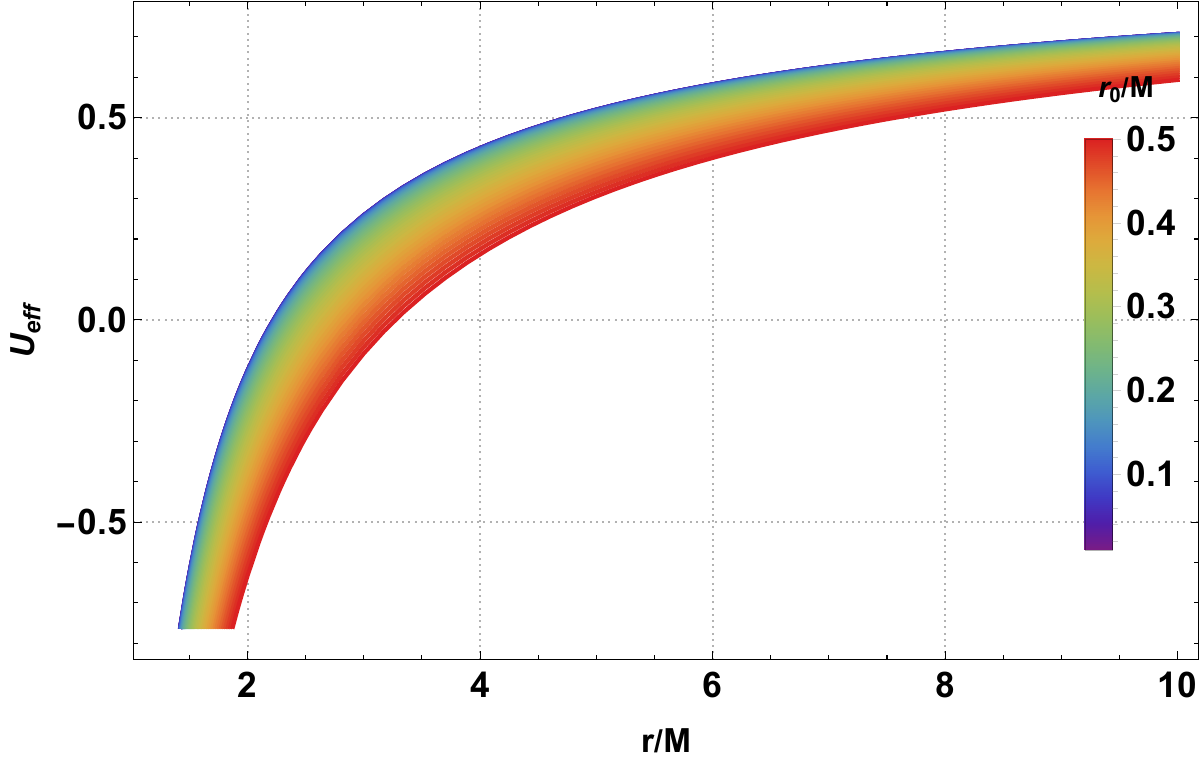}\\
    (i) $r_0/M=0.2,\;\alpha=0.1$ \hspace{4cm} (ii) $\rho_0=0.5/M^2,\;\alpha=0.1$\\[3ex]
    \includegraphics[width=0.45\linewidth]{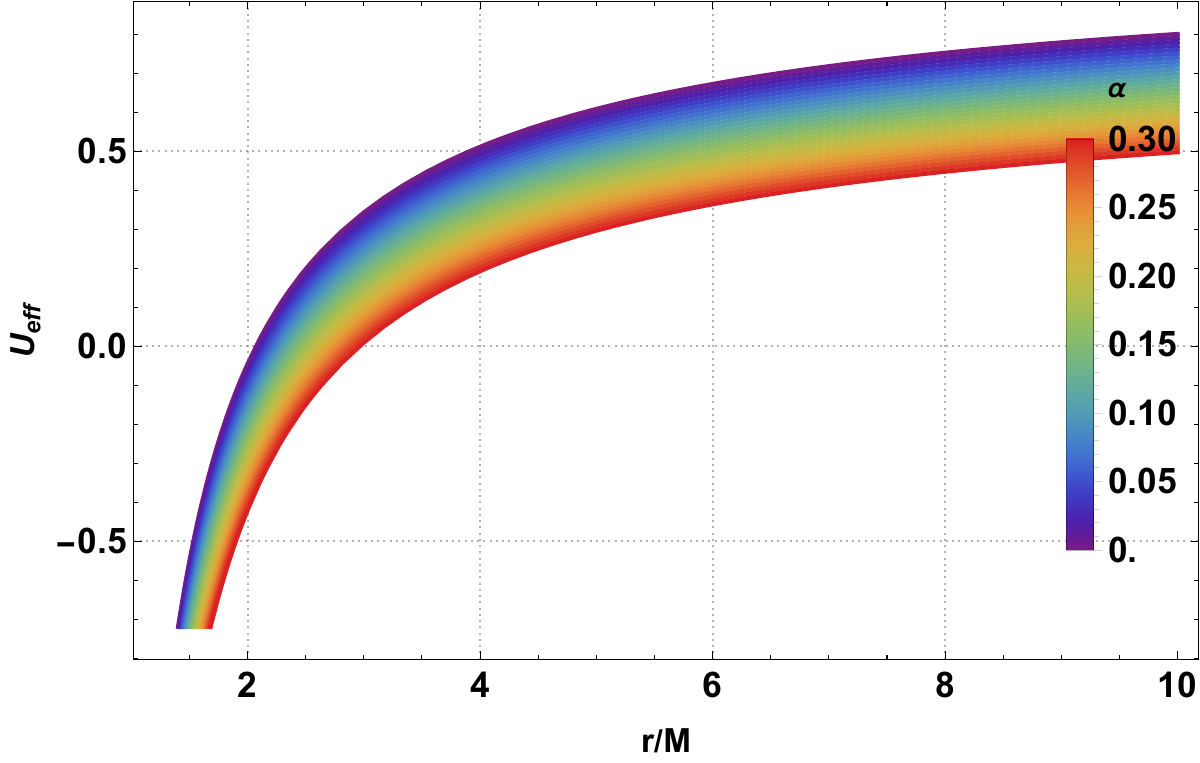}\\
    (iii) $r_0/M=0.2,\;\rho_0=0.5/M^2$
    \caption{Timelike effective potential $U_{\rm eff}(r)$ as a function of the dimensionless radial coordinate for various Plummer halo parameters $\{r_0,\,\rho_0\}$ and \CoS\ tension $\alpha$. The potential well becomes shallower as any of these parameters increases, pushing the \ISCO\ to larger radii. Here $\mathcal{L}/M=1$ and $\theta=\pi/2$.}
    \label{fig:potential-timelike}
\end{figure}

Figure~\ref{fig:potential-timelike} displays $U_{\rm eff}(r)$ in the equatorial plane ($\theta=\pi/2$) for three separate parameter scans. In panel~(i), increasing $\rho_{0}$ at fixed $r_{0}/M=0.2$ and $\alpha=0.1$ lowers both the local maximum and the local minimum of $U_{\rm eff}$, making the potential well shallower. Panel~(ii) shows a similar flattening when $r_{0}$ grows at fixed $\rho_{0}$ and $\alpha$. Panel~(iii) varies $\alpha$ at fixed $\rho_{0}$ and $r_{0}$: the entire potential curve shifts downward, with the asymptotic plateau dropping from $U_{\rm eff}(\infty)=1$ (Schwarzschild) to $1-\alpha$. In all three cases the weakening of the potential barrier implies that stable circular orbits require larger radii, i.e.\ the \ISCO\ is pushed outward.

\subsection{Specific energy and angular momentum on circular orbits}
\label{subsec:energy-circular}

For circular orbits in the equatorial plane the conditions $dr/d\lambda=0$ and $d^{2}r/d\lambda^{2}=0$ must hold simultaneously, yielding $U_{\rm eff}(r)=\mathcal{E}^{2}$ and $\partial_{r}U_{\rm eff}(r)=0$. From these two relations and the potential~(\ref{potetial-timelike}), the specific angular momentum and specific energy on a circular orbit of radius $r$ are obtained as
\begin{equation}
\mathcal{L}_{\rm sp}^{2}
=\frac{r^{3}\,A'(r)}{2\,A(r)-r\,A'(r)}
=\frac{-4\pi\rho_{0}r_{0}^{3}\,r^{2}
\!\left(-\dfrac{\arctan(r/r_{0})}{r}
+\dfrac{r_{0}}{r^{2}+r_{0}^{2}}\right)
+r_{s}\,r\;
e^{\frac{4\pi\rho_{0}r_{0}^{3}}{r}\arctan\!\left(\frac{r}{r_{0}}\right)}}
{2+4\pi\rho_{0}r_{0}^{3}
\!\left(-\dfrac{\arctan(r/r_{0})}{r}
+\dfrac{r_{0}}{r^{2}+r_{0}^{2}}\right)
-\left(\dfrac{3r_{s}}{r}+2\alpha\right)
e^{\frac{4\pi\rho_{0}r_{0}^{3}}{r}\arctan\!\left(\frac{r}{r_{0}}\right)}}\,,
\label{momentum}
\end{equation}
and
\begin{equation}
\mathcal{E}_{\rm sp}^{2}
=\frac{2\,A(r)^{2}}{2\,A(r)-r\,A'(r)}
=\frac{2\left[\exp\!\left\{-\dfrac{4\pi\rho_{0}r_{0}^{3}}{r}
\arctan\!\left(\dfrac{r}{r_{0}}\right)\right\}
-\dfrac{r_{s}}{r}-\alpha\right]^{2}}
{e^{-\frac{4\pi\rho_{0}r_{0}^{3}}{r}
\arctan\!\left(\frac{r}{r_{0}}\right)}
\left[2+4\pi\rho_{0}r_{0}^{3}
\!\left(-\dfrac{\arctan(r/r_{0})}{r}
+\dfrac{r_{0}}{r^{2}+r_{0}^{2}}\right)\right]
-\dfrac{3r_{s}}{r}-2\alpha}\,.
\label{energy}
\end{equation}
Both quantities depend on $\{\rho_{0},\,r_{0},\,\alpha,\,r_{s}\}$ and reduce to the standard Schwarzschild expressions $\mathcal{L}_{\rm sp}^{2}=Mr/(1-3M/r)$ and $\mathcal{E}_{\rm sp}^{2}=(1-2M/r)^{2}/(1-3M/r)$ in the limit $\rho_{0}\to 0$ and $\alpha\to 0$. The denominator $2A(r)-rA'(r)$ vanishes at the \PS\ radius [cf.\ Eq.~(\ref{cc6})], where both $\mathcal{L}_{\rm sp}$ and $\mathcal{E}_{\rm sp}$ diverge; this confirms that no timelike circular orbit exists inside the \PS, as expected on general grounds~\cite{Chandrasekhar1984}.

\subsection{\ISCO\ radius}
\label{subsec:ISCO}

The ISCO is the smallest radius where a test particle can orbit a black hole stably. It defines the inner edge of the accretion disk, sets the maximum efficiency of energy extraction from infalling matter, and shapes observational features such as relativistic emission lines, disk spectra, and the innermost appearance of black hole shadows.

The \ISCO\ is located where the local minimum and maximum of $U_{\rm eff}$ merge into an inflection point. This requires the three simultaneous conditions~\cite{sec4is01}
\begin{equation}
U_{\rm eff}(r)=\mathcal{E}^{2}\,,\qquad
\partial_{r}U_{\rm eff}(r)=0\,,\qquad
\partial_{r}^{2}U_{\rm eff}(r)=0\,.
\label{condition}
\end{equation}
Eliminating $\mathcal{E}$ and $\mathcal{L}$ from these three equations and using the potential~(\ref{potetial-timelike}) leads to a single condition on $A(r)$:
\begin{equation}
A(r)\,A''(r)-2\bigl[A'(r)\bigr]^{2}
+\frac{3\,A(r)\,A'(r)}{r}=0\,.
\label{condition-2}
\end{equation}
This is a transcendental equation in $r$ once the metric function~(\ref{function}) is substituted, and it must be solved numerically for each parameter set $\{\rho_{0},\,r_{0},\,\alpha\}$.

In Newtonian gravity the effective potential possesses a minimum for any nonzero angular momentum, so stable circular orbits extend down to arbitrarily small radii and the concept of an \ISCO\ does not arise. In \GR, however, the strong-field $1/r^{3}$ term in the potential creates a local maximum that eventually merges with the minimum at a critical angular momentum, defining the \ISCO~\cite{Chandrasekhar1984,Wald1984}. For the Schwarzschild \BH\ this occurs at $r_{\rm ISCO}=6M$. The Plummer \DM\ halo and \CoS\ both deepen the gravitational well at intermediate radii and lower the angular-momentum barrier, pushing the merger of extrema to larger $r$.

Table~\ref{tab:ISCO-radius} lists the numerically determined \ISCO\ radii for varying $\rho_{0}$ and $\alpha$ at fixed $r_{0}/M=0.2$. The Schwarzschild result $r_{\rm ISCO}=6M$ appears at $\rho_{0}=\alpha=0$, confirming the code. Along each row the \ISCO\ increases with $\alpha$: at $\rho_{0}=0$ it grows from $6M$ to $8.571\,M$ as $\alpha$ increases from $0$ to $0.3$, a $43\%$ enlargement. Along each column the \DM\ density produces a comparable effect: at $\alpha=0$ the \ISCO\ expands from $6M$ to $8.438\,M$ as $\rho_{0}M^{2}$ goes from $0$ to $0.5$, a $41\%$ increase. The combined effect at $(\rho_{0}M^{2},\alpha)=(0.5,\,0.3)$ pushes the \ISCO\ to $12.269\,M$, more than twice the Schwarzschild value. This substantial outward shift indicates that accretion disks around Plummer-\CoS\ \BHs\ would truncate at considerably larger radii than those around isolated Schwarzschild \BHs, reducing the radiative efficiency $\eta=1-\mathcal{E}_{\rm sp}(r_{\rm ISCO})$~\cite{sec4is01}.

\begin{table}[ht!]
\centering
\setlength{\tabcolsep}{8pt}
\renewcommand{\arraystretch}{1.4}
\begin{tabular}{|c|c|c|c|c|c|c|c|}
\hline
\rowcolor{orange!50}
\diagbox[innerwidth=2.2cm, height=1.2cm, linecolor=black]{\textbf{$\rho_0 M^2$}}{\textbf{$\alpha$}} & \textbf{0.00} & \textbf{0.05} & \textbf{0.10} & \textbf{0.15} & \textbf{0.20} & \textbf{0.25} & \textbf{0.30} \\
\hline
0.0 & 6.0000 & 6.3158 & 6.6667 & 7.0587 & 7.5000 & 8.0000 & 8.5711 \\
\hline
0.1 & 6.5498 & 6.8967 & 7.2818 & 7.7130 & 8.1975 & 8.7462 & 9.3746 \\
\hline
0.2 & 7.0609 & 7.4388 & 7.8587 & 8.3278 & 8.8550 & 9.4543 & 10.138 \\
\hline
0.3 & 7.5424 & 7.9511 & 8.4051 & 8.9129 & 9.4840 & 10.131 & 10.871 \\
\hline
0.4 & 8.0002 & 8.4394 & 8.9279 & 9.4738 & 10.088 & 10.784 & 11.580 \\
\hline
0.5 & 8.4384 & 8.9086 & 9.4311 & 10.015 & 10.672 & 11.418 & 12.269 \\
\hline
\end{tabular}
\caption{\ISCO\ radius $r_{\rm ISCO}/M$ for different values of $\rho_0 M^2$ and $\alpha$, with $r_0/M=0.2$. The Schwarzschild value $r_{\rm ISCO}=6M$ is recovered at $\rho_0=\alpha=0$. Both the Plummer \DM\ density and the \CoS\ tension enlarge the \ISCO\ monotonically.}
\label{tab:ISCO-radius}
\end{table}

\begin{figure}[ht!]
    \centering
    \includegraphics[width=0.65\linewidth]{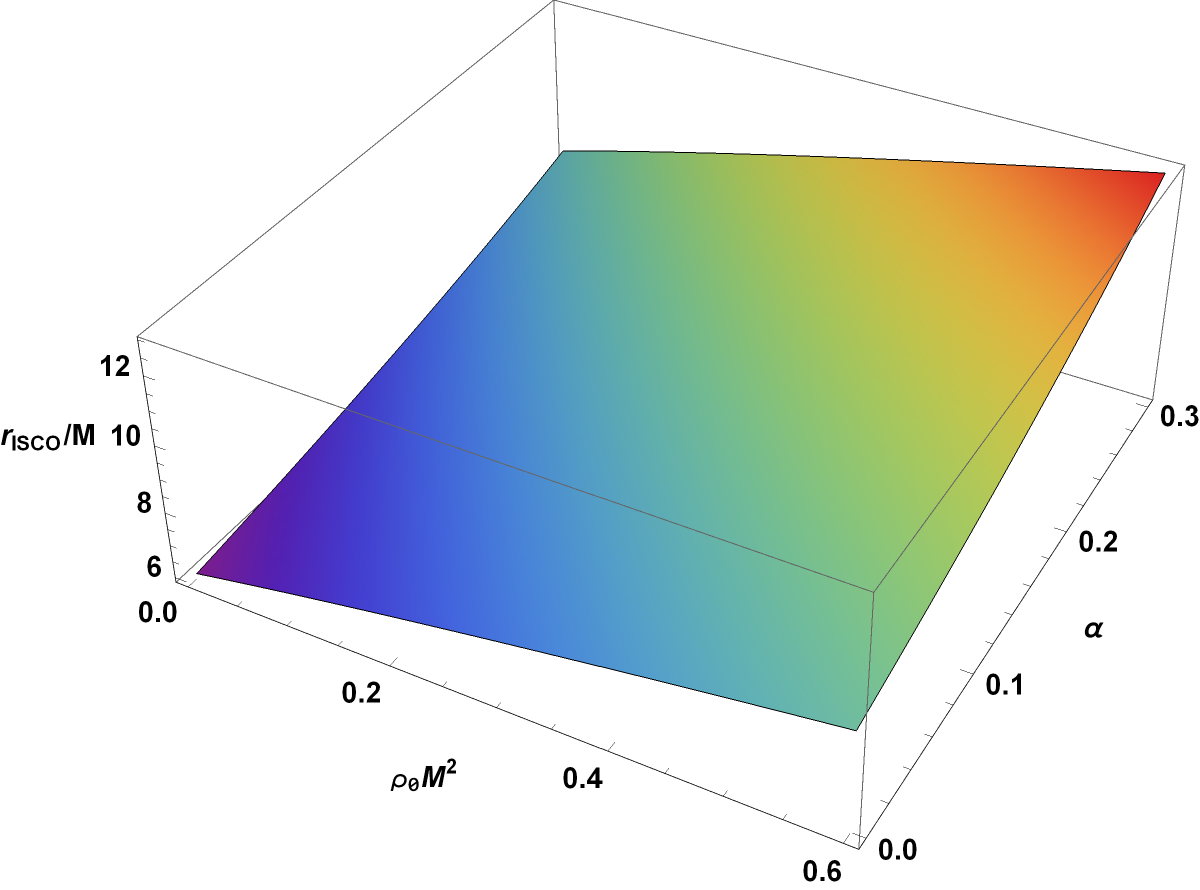}
    \caption{\ISCO\ radius $r_{\rm ISCO}/M$ as a function of $\{\rho_0,\,\alpha\}$ at fixed core radius $r_0/M=0.2$. The surface rises monotonically with both parameters, with the steeper gradient along the $\alpha$-axis.}
    \label{fig:ISCO}
\end{figure}

The three-dimensional surface plot of $r_{\rm ISCO}$ over the $(\rho_{0},\,\alpha)$ plane, shown in Fig.~\ref{fig:ISCO}, confirms these trends visually. The surface rises with a roughly linear gradient along both axes in the low-parameter regime, steepening as $\alpha$ approaches its upper bound. The gradient along $\alpha$ is slightly steeper than along $\rho_{0}$, consistent with the pattern observed for the \EH, \PS, and shadow in the preceding sections: the \CoS\ tension is the primary driver of all geometric shifts, with the Plummer halo providing a secondary, additive contribution.

\section{Scalar Perturbations, Greybody Factors, and Quasinormal Modes}
\label{sec:5}

The linear stability of a \BH\ spacetime and the spectral properties of its Hawking radiation are both governed by the effective potential experienced by perturbation fields. In this section we derive the Schr\"{o}dinger-like radial equation for a massless scalar field propagating in the Plummer-\CoS\ geometry~(\ref{metric}), examine the shape of the resulting potential barrier, compute rigorous lower bounds on the \GFs\ using the Boonserm-Visser method~\cite{Visser1998,Boonserm2008}, and determine the \QNM\ spectrum via the \WKB\ approximation~\cite{Iyer1987,Konoplya2003}.

\subsection{Klein-Gordon equation and effective potential}
\label{subsec:KG}

A massless scalar field $\Phi(t,r,\theta,\phi)$ in the background~(\ref{metric}) obeys the Klein--Gordon equation
\begin{equation}
\Box\,\Phi
=\frac{1}{\sqrt{-g}}\,\partial_{\mu}
\!\left(\sqrt{-g}\,g^{\mu\nu}\,\partial_{\nu}\Phi\right)=0\,.
\label{sf1}
\end{equation}
Exploiting the spherical symmetry and stationarity of the metric, we decompose the field as
\begin{equation}
\Phi(t,r,\theta,\phi)
=\frac{\psi(r)}{r}\,Y_{\ell m}(\theta,\phi)\,e^{-i\omega t}\,,
\label{sf2}
\end{equation}
where $Y_{\ell m}$ are the standard spherical harmonics, $\omega$ is the frequency, and $\ell=0,1,2,\ldots$ is the multipole number. Introducing the tortoise coordinate $r_{*}$ through $dr_{*}=dr/A(r)$, the radial function $\psi(r)$ satisfies the Schr\"{o}dinger-like equation
\begin{equation}
\frac{d^{2}\psi}{dr_{*}^{2}}
+\left[\omega^{2}-V_{s}(r)\right]\psi=0\,,
\label{sf3}
\end{equation}
with the scalar perturbation potential
\begin{equation}
V_{s}(r)=A(r)\left[\frac{\ell(\ell+1)}{r^{2}}
+\frac{A'(r)}{r}\right].
\label{sf4}
\end{equation}
The boundary conditions appropriate to the scattering problem are purely ingoing waves at the \EH\ ($r_{*}\to-\infty$) and purely outgoing waves at spatial infinity ($r_{*}\to+\infty$).

For the Plummer-\CoS\ metric function~(\ref{function}), the explicit form of $V_{s}$ reads
\begin{align}
V_{s}(r)=\frac{\left[\exp\!\left\{-\frac{4\pi\rho_{0}r_{0}^{3}}{r}
\tan^{-1}\!\left(\frac{r}{r_{0}}\right)\right\}-\frac{r_{s}}{r}-\alpha\right]}{r^2}\left[\ell(\ell+1)+\frac{r_{s}}{r}-\frac{4\pi\rho_{0}r_{0}^{3} e^{-\frac{4\pi\rho_{0}r_{0}^{3}}{r}\arctan(r/r_{0})}\left(r\,r_{0}-(r^{2}+r_{0}^{2})\arctan(r/r_{0})\right)}
{r (r^{2}+r_{0}^{2})}\right].
\label{sf4_explicit}
\end{align}
At the \EH\ and at spatial infinity $V_{s}$ vanishes, while it attains a positive maximum at some intermediate radius $r_{\rm peak}$ that depends on $\ell$, $\rho_{0}$, $r_{0}$, and $\alpha$. The height and width of this barrier control both the \GF\ and the \QNM\ spectrum.

\subsection{Parameter dependence of $V_{s}$}\label{subsec:Vs_plots}

\begin{figure}[ht!]
    \centering
    \includegraphics[width=0.45\linewidth]{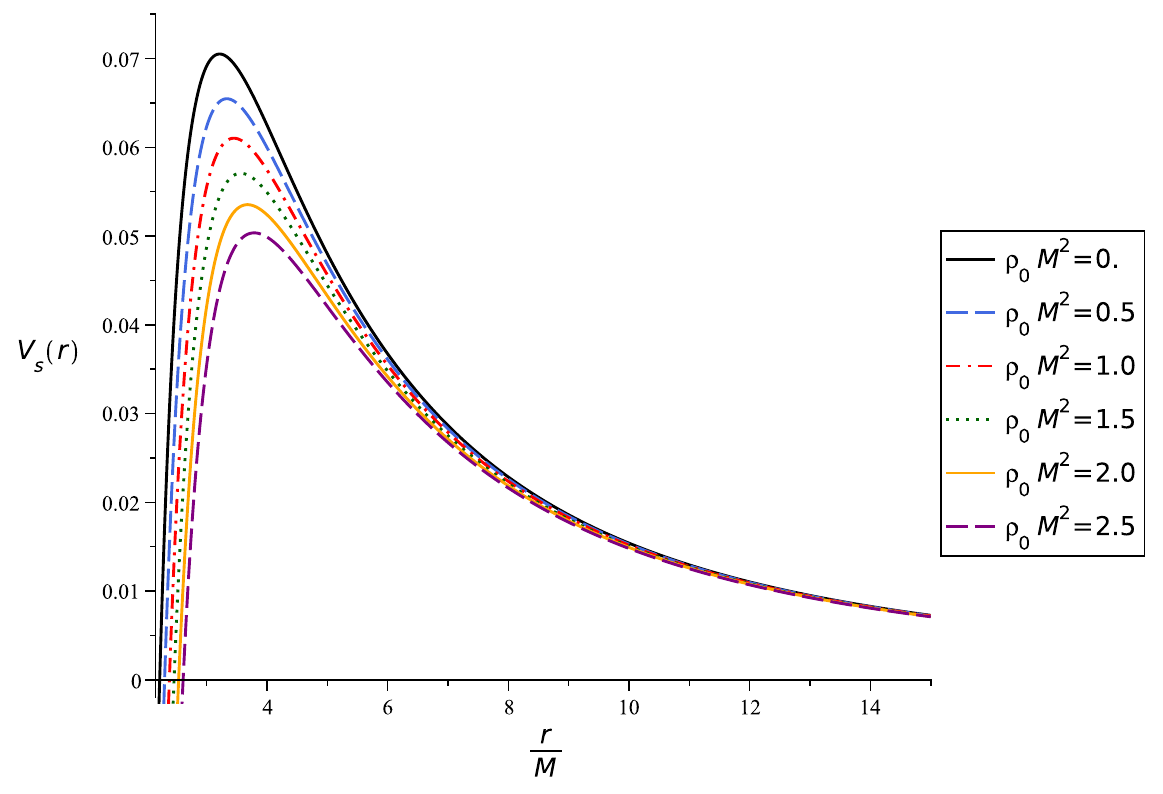}\qquad
    \includegraphics[width=0.45\linewidth]{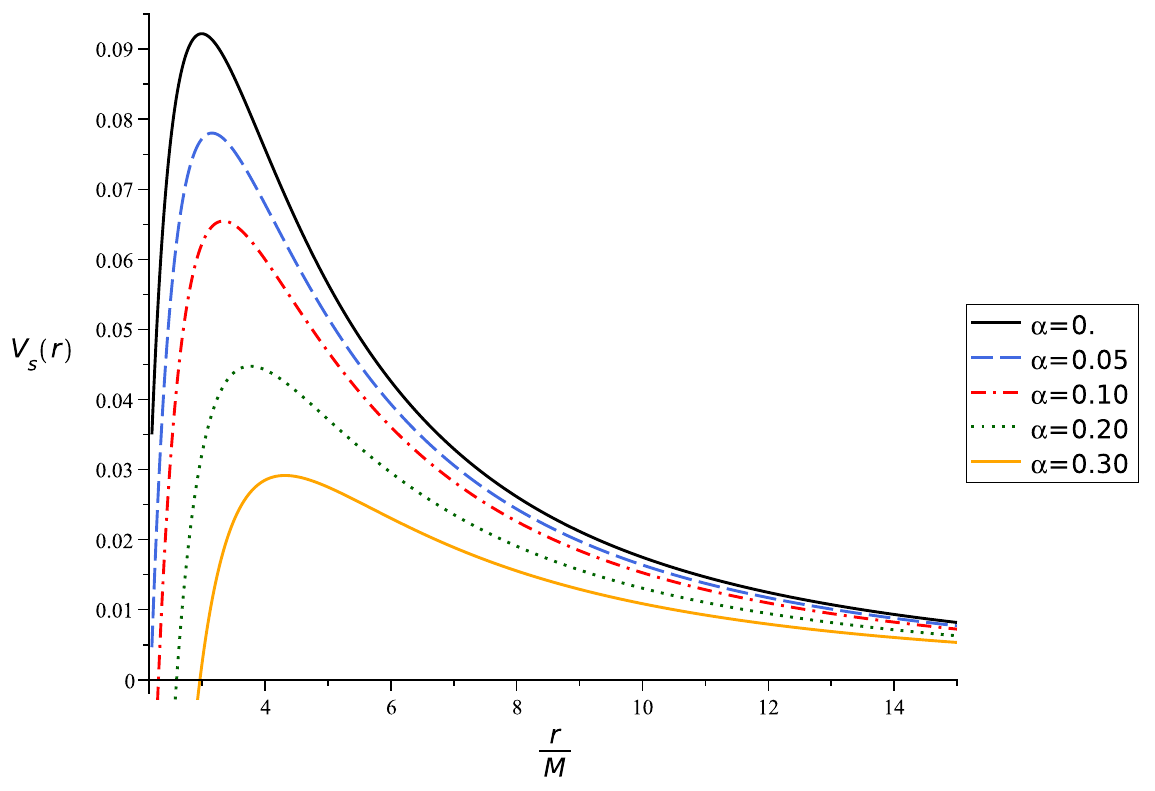}\\
    (i) Varying $\rho_0$: $r_0/M=0.2,\;\alpha=0.1,\;\ell=1$
    \hspace{1cm}
    (ii) Varying $\alpha$: $\rho_0 M^2=0.5,\;r_0/M=0.2,\;\ell=1$\\[3ex]
    \includegraphics[width=0.45\linewidth]{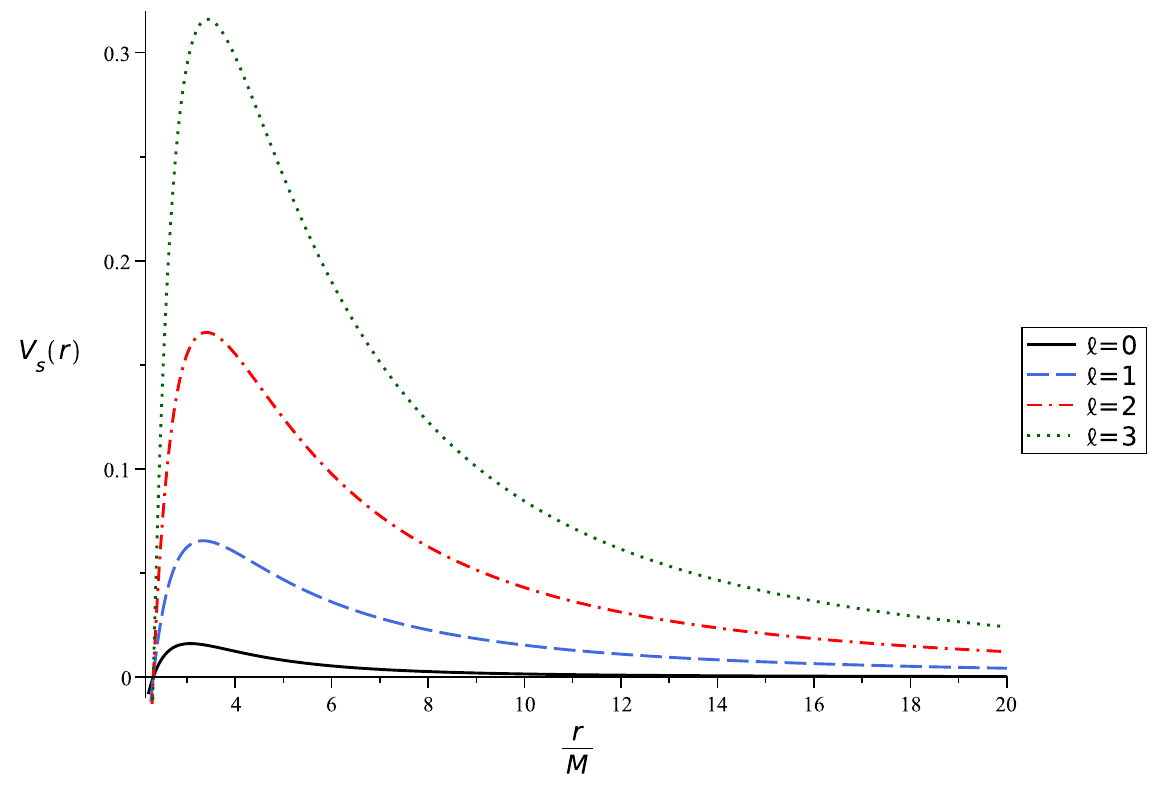}\\
    (iii) Varying $\ell$: $\rho_0 M^2=0.5,\;r_0/M=0.2,\;\alpha=0.1$
    \caption{Scalar perturbation potential $V_s(r)$ for the Plummer--\CoS\ \BH. Panel~(i): increasing $\rho_0$ lowers and broadens the barrier. Panel~(ii): increasing $\alpha$ produces a comparable suppression. Panel~(iii): higher $\ell$ modes encounter a taller, narrower barrier dominated by the centrifugal term.}
    \label{fig:Vs_panels}
\end{figure}

Figure~\ref{fig:Vs_panels} displays $V_{s}(r)$ for three parameter scans. In panel~(i) the central \DM\ density $\rho_{0}$ is varied at fixed $r_{0}/M=0.2$, $\alpha=0.1$, and $\ell=1$. The peak height decreases monotonically with increasing $\rho_{0}$: at $\rho_{0}M^{2}=0$ the peak reaches $V_{s}\approx 0.045\,M^{-2}$, while at $\rho_{0}M^{2}=2.5$ it drops to $\approx 0.030\,M^{-2}$. The peak position simultaneously shifts outward, consistent with the enlargement of the \PS\ and \EH\ reported in Secs.~\ref{sec:3} and~\ref{sec:2}. Panel~(ii) varies the \CoS\ tension $\alpha$ at fixed $\rho_{0}M^{2}=0.5$ and $\ell=1$. The suppression of the barrier is even more pronounced: increasing $\alpha$ from $0$ to $0.3$ reduces the peak by roughly $40\%$. Panel~(iii) scans the multipole number $\ell$ at fixed $\rho_{0}M^{2}=0.5$, $r_{0}/M=0.2$, and $\alpha=0.1$. As expected, higher $\ell$ modes face a taller and narrower barrier, since the centrifugal term $\ell(\ell+1)/r^{2}$ dominates at large $\ell$.

A lower potential barrier implies that a larger fraction of an incoming wave can tunnel through to the \EH, leading to a higher \GF. Conversely, a taller barrier reflects more of the incident radiation back to infinity.

\subsection{Greybody factor bounds}
\label{subsec:GF_bound}

The \GF\ $|\mathcal{T}_{\ell}(\omega)|^{2}$ gives the probability that a scalar wave of frequency $\omega$ and angular momentum $\ell$ tunnels through the potential barrier $V_{s}$ and reaches the \EH. A rigorous lower bound on $|\mathcal{T}_{\ell}|^{2}$ was derived by Visser~\cite{Visser1998} and refined by Boonserm et al.~\cite{Boonserm2008}:
\begin{equation}
|\mathcal{T}_{\ell}|^{2}
\;\ge\;
\operatorname{sech}^{2}\!\left(
\frac{1}{2\omega}\int_{r_{h}}^{\infty}
\frac{|V_{s}(r)|}{A(r)}\,dr
\right).
\label{eq:GF_bound}
\end{equation}
The integral is evaluated numerically using a midpoint-rule quadrature over $r\in[r_{h}+\epsilon,\,50\,M]$, which gives convergence to four significant figures.

\begin{figure}[ht!]
    \centering
    \includegraphics[width=0.75\linewidth]{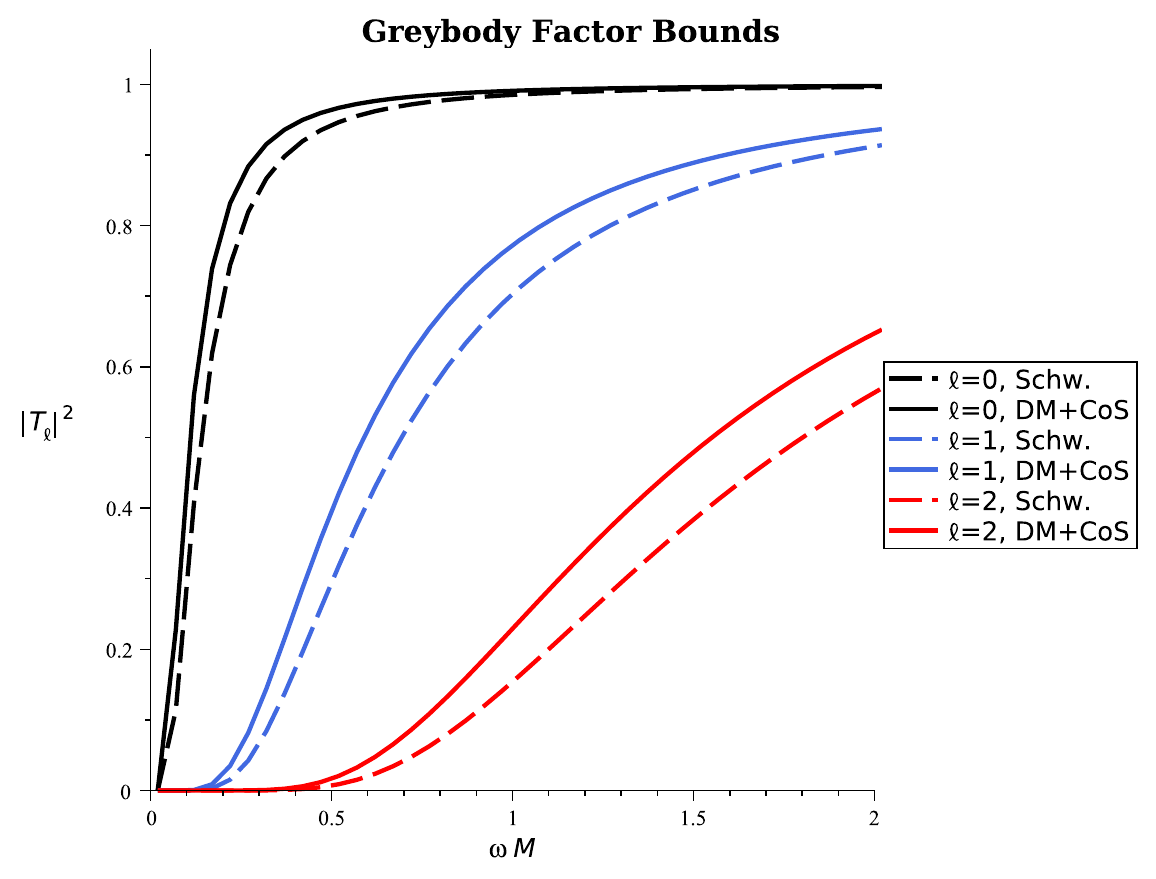}
    \caption{\GF\ bounds $|\mathcal{T}_{\ell}|^{2}$ vs $\omega M$ for $\ell=0$ (black), $\ell=1$ (blue), $\ell=2$ (red). Dashed: Schwarzschild; solid: Plummer-\CoS\ with $\rho_0 M^2=0.5$, $\alpha=0.1$, $r_0/M=0.2$. The \DM\ halo and \CoS\ lower the potential barrier, enhancing transmission for all modes.}
    \label{fig:GF1}
\end{figure}

Figure~\ref{fig:GF1} presents $|\mathcal{T}_{\ell}|^{2}$ as a function of $\omega M$ for $\ell=0,\,1,\,2$, comparing the Schwarzschild baseline (dashed curves) with the Plummer-\CoS\ configuration at $\rho_{0}M^{2}=0.5$ and $\alpha=0.1$ (solid curves). For each $\ell$, the Plummer-\CoS\ \GF\ lies above the Schwarzschild result across the entire frequency range, reflecting the lower potential barrier identified in Fig.~\ref{fig:Vs_panels}. The enhancement is most visible for $\ell=0$: at $\omega M=0.2$ the \GF\ rises from $\approx 0.45$ (Schwarzschild) to $\approx 0.55$ (Plummer-\CoS), a $\sim 22\%$ increase.

\begin{figure}[ht!]
    \centering
    \includegraphics[width=0.65\linewidth]{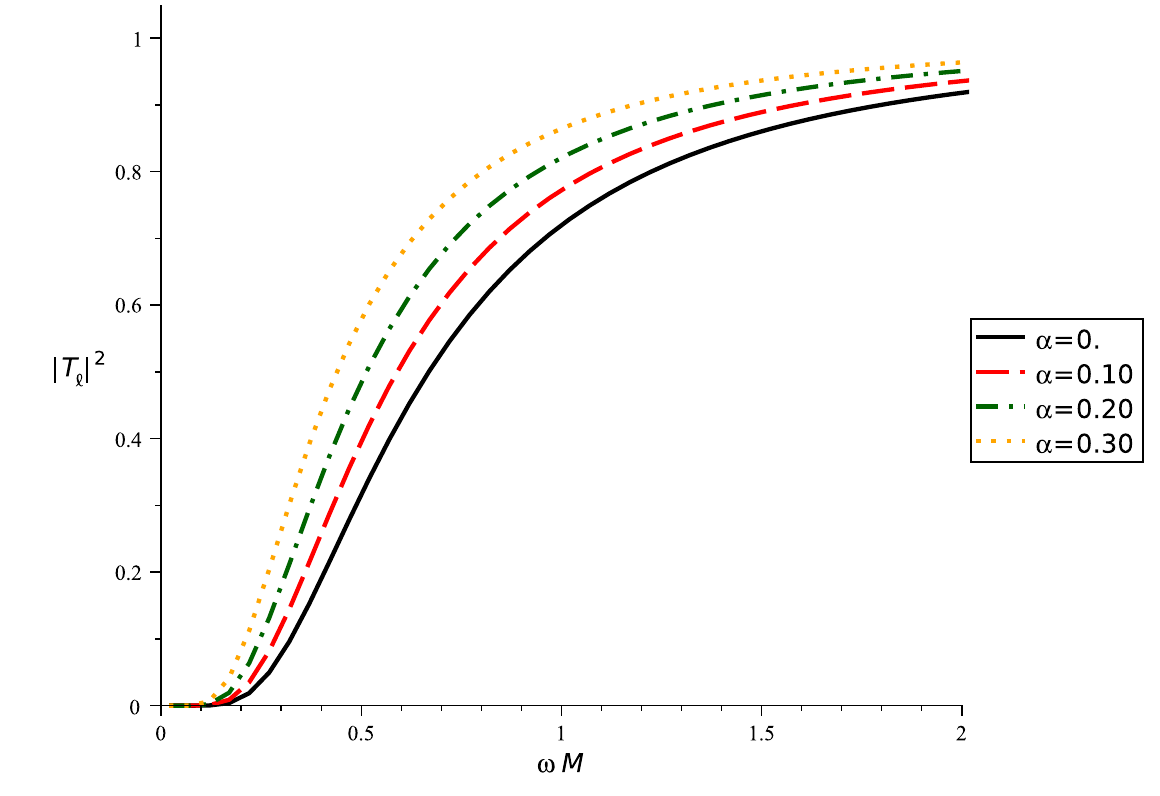}\\
    (i) Varying $\alpha$: $\rho_0 M^2=0.5,\;r_0/M=0.2,\;\ell=1$\\[3ex]
    \includegraphics[width=0.65\linewidth]{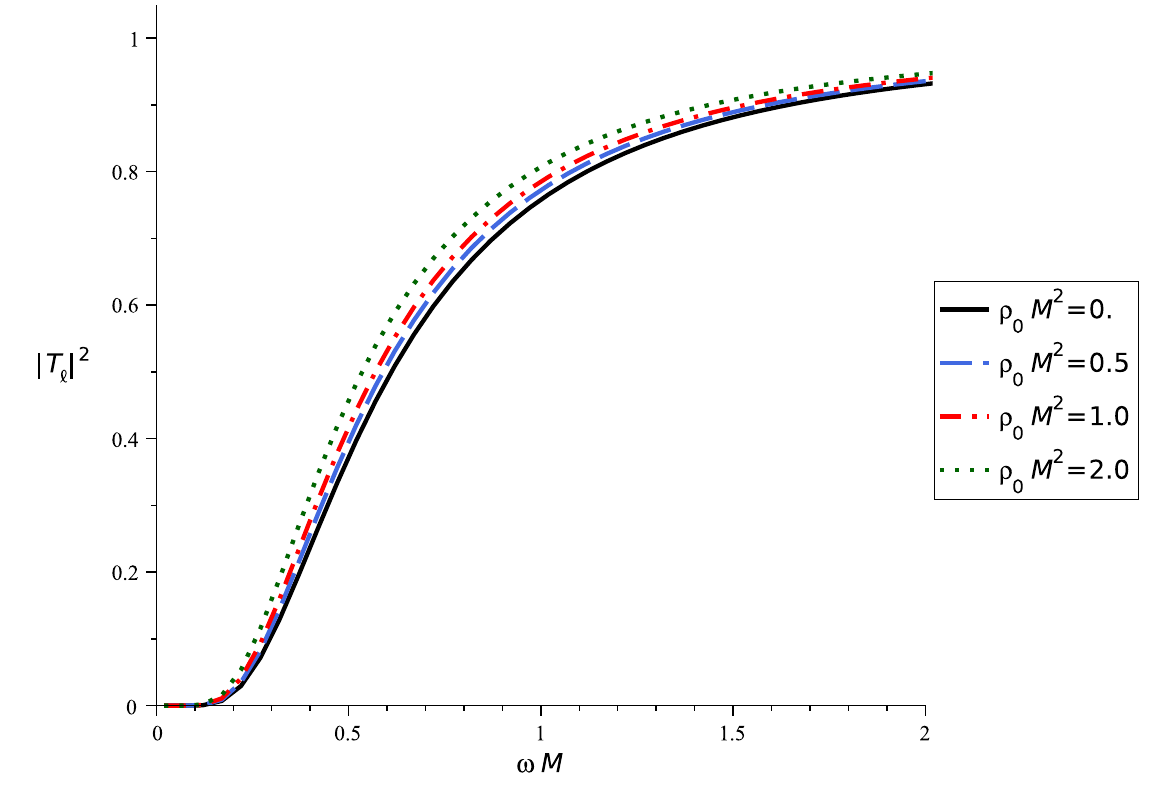}\\
    (ii) Varying $\rho_0$: $\alpha=0.1,\;r_0/M=0.2,\;\ell=1$
    \caption{\GF\ bound $|\mathcal{T}_{1}|^{2}$ vs $\omega M$. Panel~(i): varying $\alpha$ at $\rho_0 M^2=0.5$. Panel~(ii): varying $\rho_0$ at $\alpha=0.1$. In both panels $r_0/M=0.2$.}
    \label{fig:GF23}
\end{figure}

The separate roles of the \DM\ halo and the \CoS\ are disentangled in Fig.~\ref{fig:GF23}. In panel~(i) the \CoS\ tension $\alpha$ is varied at fixed $\rho_{0}M^{2}=0.5$ and $\ell=1$: larger $\alpha$ progressively raises the \GF\ curve. In panel~(ii) the \DM\ density $\rho_{0}$ is scanned at fixed $\alpha=0.1$ and $\ell=1$: higher $\rho_{0}$ also enhances transmission, though less dramatically than the \CoS\ variation.

\subsection{Hawking emission spectrum}
\label{subsec:emission}

The differential energy emission rate for scalar (bosonic) radiation from the \BH\ is given by~\cite{sec5is01,sec5is02}
\begin{equation}
\frac{d^{2}E}{dt\,d\omega}
=\frac{1}{2\pi}\sum_{\ell=0}^{\infty}(2\ell+1)\,
\frac{\omega\,|\mathcal{T}_{\ell}(\omega)|^{2}}
{e^{\omega/T_{H}}-1}\,,
\label{eq:emission}
\end{equation}
where $T_{H}$ is the Hawking temperature derived in Sec.~\ref{sec:6}. The net effect of the Plummer \DM\ halo and \CoS\ on the emission spectrum is twofold: the enhanced \GFs\ increase the overall transmission \cite{SakalliKanzi2022Review,Sakalli2016PRD,Sakalli2018GRG,Sakalli2016ApSS}, while the reduced $T_{H}$ suppresses the thermal factor. These two effects compete, and the balance depends on the specific values of $\rho_{0}$ and $\alpha$.

\begin{figure}[ht!]
    \centering
    \includegraphics[width=0.75\linewidth]{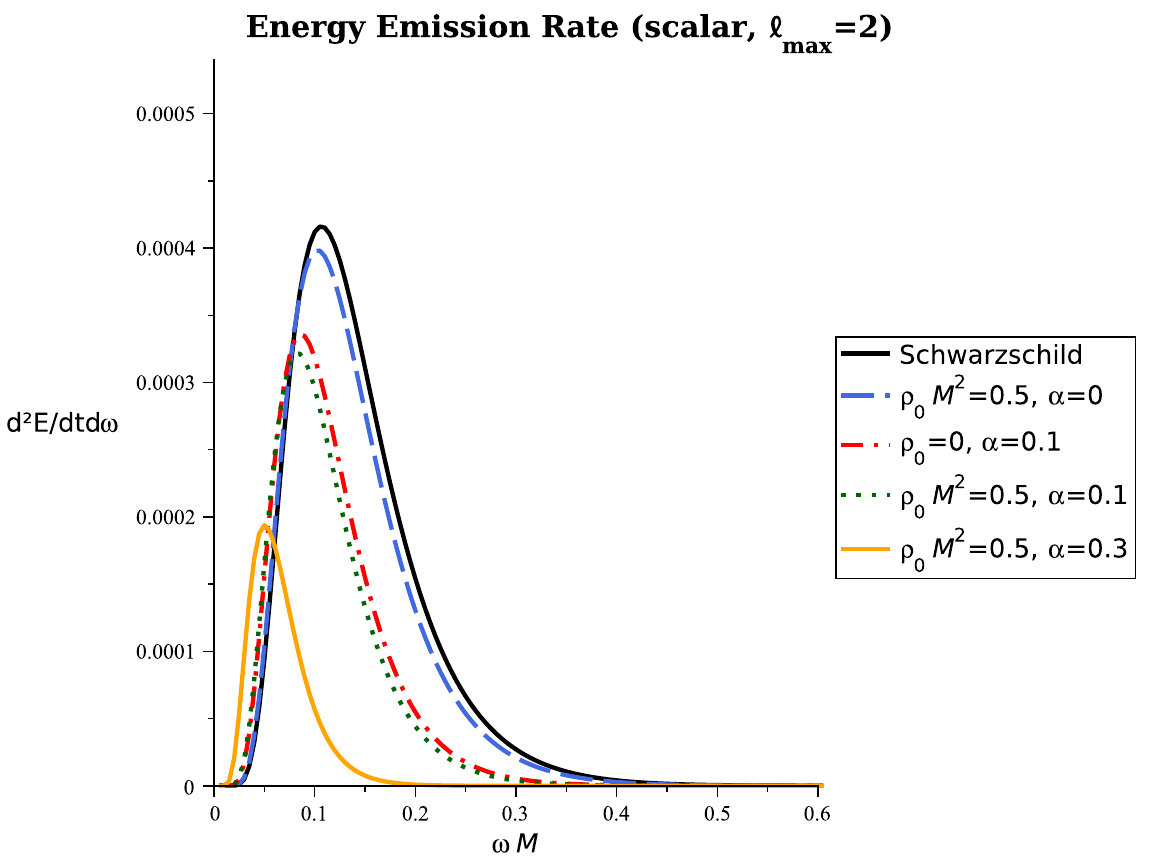}
    \caption{Scalar energy emission rate $d^{2}E/(dt\,d\omega)$ vs $\omega M$, summed over $\ell=0,1,2$. The Schwarzschild baseline (solid black) is compared with four Plummer-\CoS\ configurations. Increasing $\rho_0$ or $\alpha$ reduces the peak emission. Parameters: $r_0/M=0.2$, $M=1$.}
    \label{fig:emission}
\end{figure}

Figure~\ref{fig:emission} shows the energy emission rate computed by summing over $\ell=0,1,2$ for five configurations. The Schwarzschild curve (solid black) peaks near $\omega M\approx 0.10$ and decays exponentially for $\omega M\gtrsim 0.3$. Adding the Plummer halo alone (blue dashed) slightly reduces the peak height because the lower $T_{H}$ weakens the thermal factor. The \CoS\ ($\alpha=0.1$, red dash-dotted) produces a more pronounced suppression and shifts the peak to lower frequencies. At $\alpha=0.3$ (orange solid) the peak drops further and narrows, indicating that the thermal suppression overtakes the \GF\ enhancement. The overall trend is that the Plummer-\CoS\ \BH\ radiates less total power than a Schwarzschild \BH\ of the same mass.

\subsection{Quasinormal modes via the WKB method}
\label{subsec:QNMs}

The \QNMs\ of a \BH\ are the complex eigenfrequencies $\omega_{\rm QNM}=\omega_{R}+i\,\omega_{I}$ obtained by imposing purely ingoing boundary conditions at the \EH\ and purely outgoing conditions at infinity~\cite{sec2is06,Iyer1987,Konoplya2003}. The real part $\omega_{R}$ determines the oscillation frequency of the ringdown signal, and the imaginary part $\omega_{I}<0$ controls the damping time $\tau=1/|\omega_{I}|$.

We employ the \WKB\ approximation developed by Iyer and Will~\cite{Iyer1987}. At first order, the \QNM\ frequency satisfies
\begin{equation}
\omega^{2}=V_{0}-i\left(n+\tfrac{1}{2}\right)\sqrt{-2V_{0}''}\,,
\label{eq:WKB}
\end{equation}
where $V_{0}\equiv V_{s}(r_{\rm peak})$ is the potential maximum, $V_{0}''\equiv d^{2}V_{s}/dr_{*}^{2}|_{r_{\rm peak}}<0$ is its second tortoise-coordinate derivative at the peak, and $n=0,1,2,\ldots$ is the overtone number. At the peak, where $dV_{s}/dr=0$, the conversion simplifies to $V_{0}''=A(r_{\rm peak})^{2}\,d^{2}V_{s}/dr^{2}|_{r_{\rm peak}}$. The method is most accurate for $\ell\gg n$; for $\ell\ge 2$ and $n=0$ it reproduces known Schwarzschild values to within a few percent~\cite{sec2is06}.

The numerical results are compiled in Table~\ref{tab:QNM} for $\ell=1,2,3$ and overtones $n=0,1$ across six parameter configurations. In the Schwarzschild limit ($\rho_{0}=\alpha=0$) the code yields $\omega M\approx 0.5063 - 0.0961\,i$ for $\ell=2$, $n=0$, which agrees with the known value $0.4836 - 0.0968\,i$~\cite{sec2is06} to within the accuracy expected from first-order \WKB; the agreement improves for higher $\ell$.

Both $\omega_{R}$ and $|\omega_{I}|$ decrease monotonically as either $\rho_{0}$ or $\alpha$ increases. For $\ell=2$, $n=0$ the oscillation frequency drops from $\omega_{R}M = 0.5063$ (Schwarzschild) to $0.2799$ (Plummer-\CoS\ with $\alpha=0.3$), a $45\%$ reduction, while $|\omega_{I}|$ decreases by $53\%$ over the same range. These shifts reflect the lowering of the potential barrier: a shallower barrier supports a lower-frequency, longer-lived trapped mode. The quality factor $\mathcal{Q}=|\omega_{R}/(2\omega_{I})|$ increases with both $\rho_{0}$ and $\alpha$-for $\ell=2$, $n=0$ it rises from $2.63$ to $3.09$ at $\alpha=0.3$-meaning the ringdown signal becomes more monochromatic in the presence of \DM\ and \CoS. Comparing the Plummer-only row ($\rho_{0}M^{2}=0.5$, $\alpha=0$) with the \CoS-only row ($\rho_{0}=0$, $\alpha=0.1$), the latter produces a larger shift in both $\omega_{R}$ and $\omega_{I}$, consistent with the parameter hierarchy established in all preceding sections.

\begin{center}
\setlength{\tabcolsep}{12pt}
\renewcommand{\arraystretch}{1.6}
\begin{longtable}{ccccccc}
\hline\hline
\rowcolor{orange!50}
$\rho_0 M^2$ & $\alpha$ & $\ell$ & $n$ & $\omega_R M$ & $\omega_I M$ & $\mathcal{Q}$ \\
\hline
\endfirsthead
\multicolumn{7}{c}{\tablename\ \thetable\ -- \textit{Continued from previous page}} \\
\hline
\rowcolor{orange!50}
$\rho_0 M^2$ & $\alpha$ & $\ell$ & $n$ & $\omega_R M$ & $\omega_I M$ & $\mathcal{Q}$ \\
\hline
\endhead
\hline
\multicolumn{7}{r}{\textit{Continued on next page}} \\
\endfoot
\hline\hline
\endlastfoot
0.0 & 0.00 & 1 & 0 & 0.329434 & $-0.096256$ & 1.7112 \\
0.5 & 0.00 & 1 & 0 & 0.317466 & $-0.092662$ & 1.7130 \\
0.0 & 0.10 & 1 & 0 & 0.276725 & $-0.077942$ & 1.7752 \\
0.5 & 0.10 & 1 & 0 & 0.266631 & $-0.075027$ & 1.7769 \\
0.5 & 0.20 & 1 & 0 & 0.219688 & $-0.059259$ & 1.8536 \\
0.5 & 0.30 & 1 & 0 & 0.176687 & $-0.045355$ & 1.9478 \\
\hline
0.0 & 0.00 & 1 & 1 & 0.396143 & $-0.240141$ & 0.8248 \\
0.5 & 0.00 & 1 & 1 & 0.381661 & $-0.231228$ & 0.8253 \\
0.0 & 0.10 & 1 & 1 & 0.330053 & $-0.196045$ & 0.8418 \\
0.5 & 0.10 & 1 & 1 & 0.317948 & $-0.188753$ & 0.8422 \\
0.5 & 0.20 & 1 & 1 & 0.259589 & $-0.150450$ & 0.8627 \\
0.5 & 0.30 & 1 & 1 & 0.206633 & $-0.116345$ & 0.8880 \\
\hline
0.0 & 0.00 & 2 & 0 & 0.506317 & $-0.096123$ & 2.6337 \\
0.5 & 0.00 & 2 & 0 & 0.488004 & $-0.092537$ & 2.6368 \\
0.0 & 0.10 & 2 & 0 & 0.429399 & $-0.077861$ & 2.7575 \\
0.5 & 0.10 & 2 & 0 & 0.413796 & $-0.074951$ & 2.7604 \\
0.5 & 0.20 & 2 & 0 & 0.344365 & $-0.059219$ & 2.9076 \\
0.5 & 0.30 & 2 & 0 & 0.279868 & $-0.045338$ & 3.0864 \\
\hline
0.0 & 0.00 & 2 & 1 & 0.561096 & $-0.260216$ & 1.0781 \\
0.5 & 0.00 & 2 & 1 & 0.540705 & $-0.250554$ & 1.0790 \\
0.0 & 0.10 & 2 & 1 & 0.472611 & $-0.212227$ & 1.1135 \\
0.5 & 0.10 & 2 & 1 & 0.455367 & $-0.204327$ & 1.1143 \\
0.5 & 0.20 & 2 & 1 & 0.376200 & $-0.162622$ & 1.1567 \\
0.5 & 0.30 & 2 & 1 & 0.303344 & $-0.125488$ & 1.2087 \\
\hline
0.0 & 0.00 & 3 & 0 & 0.691728 & $-0.096148$ & 3.5972 \\
0.5 & 0.00 & 3 & 0 & 0.666746 & $-0.092562$ & 3.6016 \\
0.0 & 0.10 & 3 & 0 & 0.588498 & $-0.077884$ & 3.7780 \\
0.5 & 0.10 & 3 & 0 & 0.567139 & $-0.074973$ & 3.7823 \\
0.5 & 0.20 & 3 & 0 & 0.473494 & $-0.059237$ & 3.9966 \\
0.5 & 0.30 & 3 & 0 & 0.386073 & $-0.045353$ & 4.2563 \\
\hline
0.0 & 0.00 & 3 & 1 & 0.736622 & $-0.270865$ & 1.3598 \\
0.5 & 0.00 & 3 & 1 & 0.709927 & $-0.260795$ & 1.3611 \\
0.0 & 0.10 & 3 & 1 & 0.623605 & $-0.220498$ & 1.4141 \\
0.5 & 0.10 & 3 & 1 & 0.600906 & $-0.212281$ & 1.4154 \\
0.5 & 0.20 & 3 & 1 & 0.499110 & $-0.168591$ & 1.4802 \\
0.5 & 0.30 & 3 & 1 & 0.404768 & $-0.129775$ & 1.5595 \\
\caption{Scalar \QNM\ frequencies $\omega_{\rm QNM}M=\omega_R M+i\,\omega_I M$ and quality factor $\mathcal{Q}=|\omega_R/(2\omega_I)|$ for the Plummer-\CoS\ \BH, computed via 1st-order \WKB. Rows are grouped by $(\ell,n)$. Here $r_0/M=0.2$ and $M=1$.}
\label{tab:QNM}
\end{longtable}
\end{center}

\section{Thermodynamics}\label{sec:6}

In this section we derive the thermodynamic quantities-mass, Hawking temperature, entropy, heat capacity, and Gibbs free energy \cite{Bekenstein1973,Bekenstein1974,Hawking1975,Hawking1976,Davies1989} for the Plummer-\CoS\ \BH\ and analyze its local and global stability. Unlike charged or rotating \BHs, the present solution possesses a single, non-degenerate \EH\ (Sec.~\ref{sec:2}), and its thermodynamic behavior turns out to be qualitatively similar to the Schwarzschild case, with modifications controlled by the \DM\ density $\rho_{0}$ and the \CoS\ tension $\alpha$.

\subsection{BH mass and Hawking temperature}
\label{subsec:TH}

The \BH\ mass expressed in terms of the \EH\ radius $r_{h}$ follows from $A(r_{h})=0$ (with $r_{s}=2M$):
\begin{equation}
M=\frac{r_{h}}{2}\left[
\exp\!\left\{-\frac{4\pi\rho_{0}r_{0}^{3}}{r_{h}}
\tan^{-1}\!\left(\frac{r_{h}}{r_{0}}\right)\right\}
-\alpha\right].
\label{eq:mass}
\end{equation}
The Hawking temperature is obtained from the surface gravity $\kappa=A'(r_{h})/2$:
\begin{equation}
T_{H}=\frac{A'(r_{h})}{4\pi}\,.
\label{eq:TH_def}
\end{equation}
Defining the shorthand
\begin{equation}
\mathcal{C}\equiv\exp\!\left(-\frac{4\pi\rho_{0}r_{0}^{3}}{r_{h}}
\tan^{-1}\!\frac{r_{h}}{r_{0}}\right),\qquad
\mathcal{D}\equiv 4\pi\rho_{0}r_{0}^{3}
\left(\frac{r_{0}}{r_{h}^{2}+r_{0}^{2}}
-\frac{\tan^{-1}(r_{h}/r_{0})}{r_{h}}\right),
\label{eq:CD_def}
\end{equation}
and using the horizon condition $A(r_{h})=0$ to eliminate the $-r_{s}/r_{h}$ term, the temperature takes the form
\begin{equation}
T_{H}=\frac{1}{4\pi r_{h}}
\left[\mathcal{C}\left(1+\mathcal{D}\right)-\alpha\right].
\label{eq:TH_explicit}
\end{equation}
In the Schwarzschild limit ($\rho_{0}=\alpha=0$) we have $\mathcal{C}=1$ and $\mathcal{D}=0$, giving $T_{H}=1/(4\pi r_{h})=1/(8\pi M)$, as expected. The positivity of $T_{H}$ for all $\alpha<1$ and physical $\rho_{0}$ is guaranteed by the horizon condition.

The numerical data in Table~\ref{tab:thermo_data} confirm that $T_{H}$ is positive and monotonically decreasing with $r_{h}$ for all parameter combinations studied. At the Schwarzschild horizon ($r_{h}\approx 2M$, $\rho_{0}=\alpha=0$), the temperature is $T_{H}M\approx 0.0379$; adding \DM\ ($\rho_{0}M^{2}=0.5$) slightly reduces this to $T_{H}M\approx 0.0365$, while adding the \CoS\ ($\alpha=0.1$) produces a more visible drop to $T_{H}M\approx 0.0308$. For $\alpha=0.5$ the horizon shifts to $r_{h}\approx 4M$ and the temperature falls to $T_{H}M\approx 0.0097$.

\begin{figure}[ht!]
    \centering
    \includegraphics[width=0.75\linewidth]{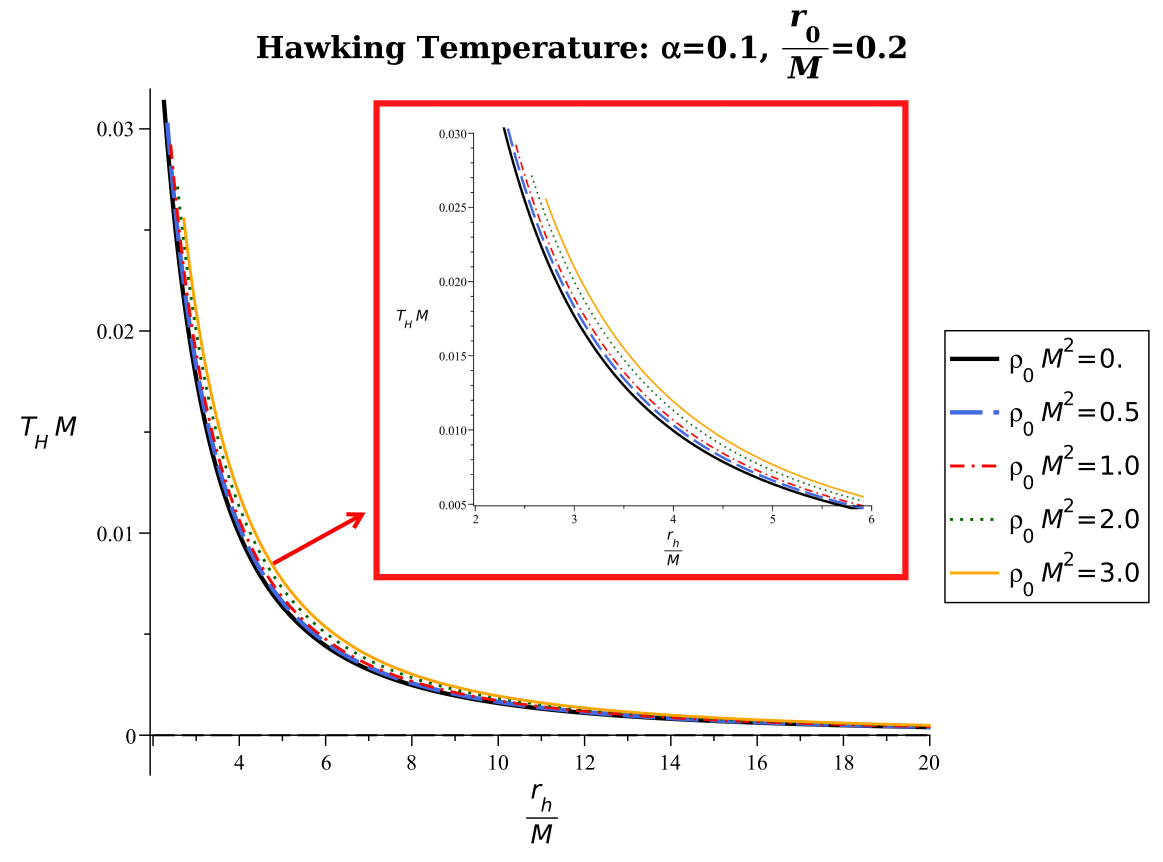}
    \caption{Hawking temperature $T_H M$ vs $r_h/M$ for varying $\rho_0$ at fixed $\alpha=0.1$ and $r_0/M=0.2$. Inset (red box): zoomed near-horizon region $r_h/M\in[2,6]$ showing the separation between curves. Higher $\rho_0$ suppresses the temperature at each $r_h$.}
    \label{fig:TH}
\end{figure}

Figure~\ref{fig:TH} shows the Hawking temperature as a function of the \EH\ radius for varying $\rho_{0}$ at fixed $\alpha=0.1$ and $r_{0}/M=0.2$. All curves follow the expected $T_{H}\sim 1/r_{h}$ decay at large $r_{h}$, where the Schwarzschild-like behavior dominates. The curves separate most visibly near the horizon: higher $\rho_{0}$ reduces $T_{H}$ at each $r_{h}$ because the enclosed \DM\ mass shifts the horizon outward (cf.\ Table~\ref{tab:horizon-longtable}), effectively producing a more massive-and therefore colder-\BH\ at the same geometric radius. The red-boxed inset zooms into the near-horizon region $r_{h}/M\in[2,6]$, where the ordering of the five curves is clearly resolved.

\subsection{Entropy, heat capacity, and local stability}
\label{subsec:entropy}

The Bekenstein--Hawking entropy follows from the area law~\cite{sec6is01}:
\begin{equation}
S=\frac{\mathcal{A}}{4}=\pi r_{h}^{2}\,,
\label{eq:entropy}
\end{equation}
where $\mathcal{A}=4\pi r_{h}^{2}$ is the area of the two-sphere at the \EH. This expression retains its standard form because neither the Plummer \DM\ profile nor the \CoS\ modifies the angular part of the metric; only the lapse function $A(r)$ is altered. However, since $r_{h}$ itself depends on $\rho_{0}$ and $\alpha$ (Sec.~\ref{subsec:horizon}), the entropy at fixed \BH\ mass is indirectly affected by both parameters: a larger $r_{h}$ at given $M$ means a larger entropy.

The canonical heat capacity at constant parameters is~\cite{sec6is02}
\begin{equation}
C_{H}=\frac{\partial M}{\partial T_{H}}
=\frac{\partial M/\partial r_{h}}
{\partial T_{H}/\partial r_{h}}\,.
\label{eq:CH_def}
\end{equation}
Using the mass~(\ref{eq:mass}) and the temperature~(\ref{eq:TH_explicit}), the heat capacity can be written as
\begin{equation}
C_{H}=2\pi r_{h}^{2}\,
\frac{\mathcal{C}\left(1+\mathcal{D}\right)-\alpha}
{-r_{h}\,\mathcal{C}\left(\gamma'\mathcal{D}+\mathcal{D}'\right)
-\alpha-\mathcal{C}\mathcal{D}}\,,
\label{eq:CH_explicit}
\end{equation}
where
\begin{equation}
\gamma'=-4\pi\rho_{0}r_{0}^{3}\,
\frac{r_{h}\,r_{0}-(r_{h}^{2}+r_{0}^{2})\tan^{-1}(r_{h}/r_{0})}
{r_{h}^{2}(r_{h}^{2}+r_{0}^{2})}\,,\qquad
\mathcal{D}'=4\pi\rho_{0}r_{0}^{3}
\left[\gamma'/r_{h}^{2}
-\frac{2r_{0}\,r_{h}}{(r_{h}^{2}+r_{0}^{2})^{2}}\right].
\label{eq:gamma_Dprime}
\end{equation}
A positive $C_{H}$ signals local thermodynamic stability, while $C_{H}<0$ indicates instability.

\begin{figure}[ht!]
    \centering
    \includegraphics[width=0.65\linewidth]{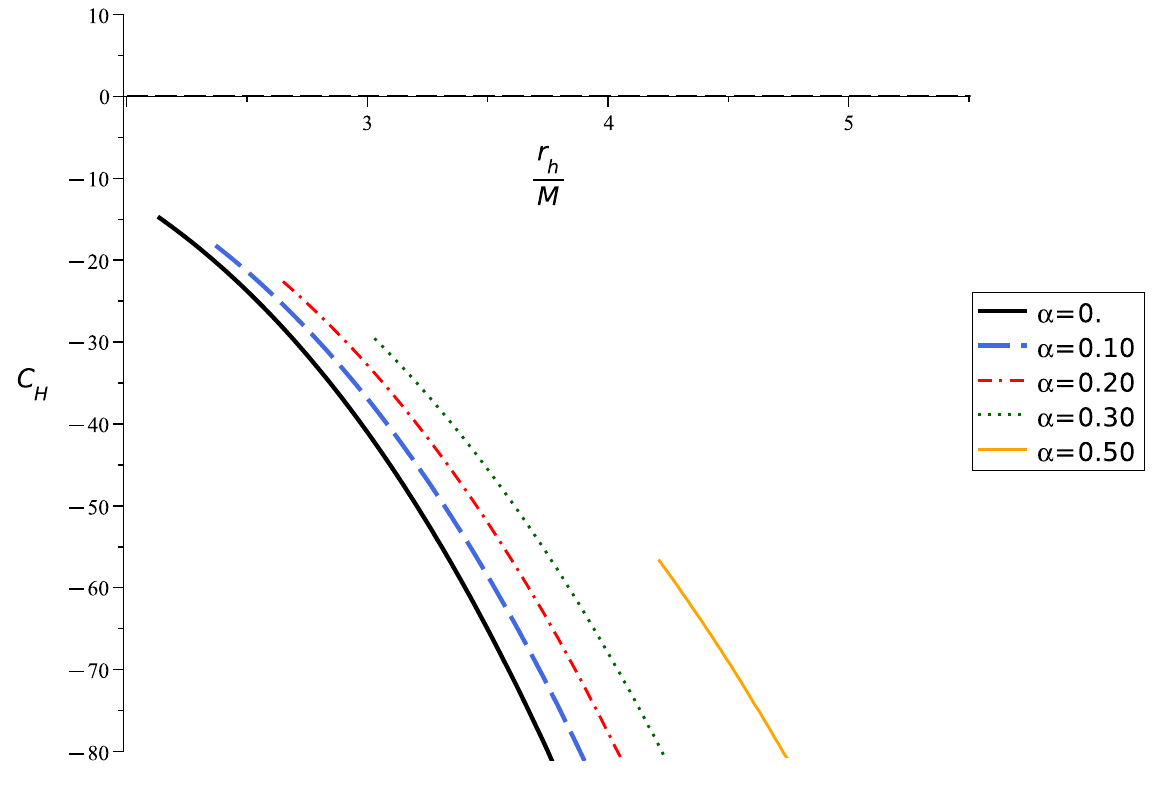}\\
    (i) Varying $\alpha$: $\rho_0 M^2=0.5,\;r_0/M=0.2$\\[3ex]
    \includegraphics[width=0.65\linewidth]{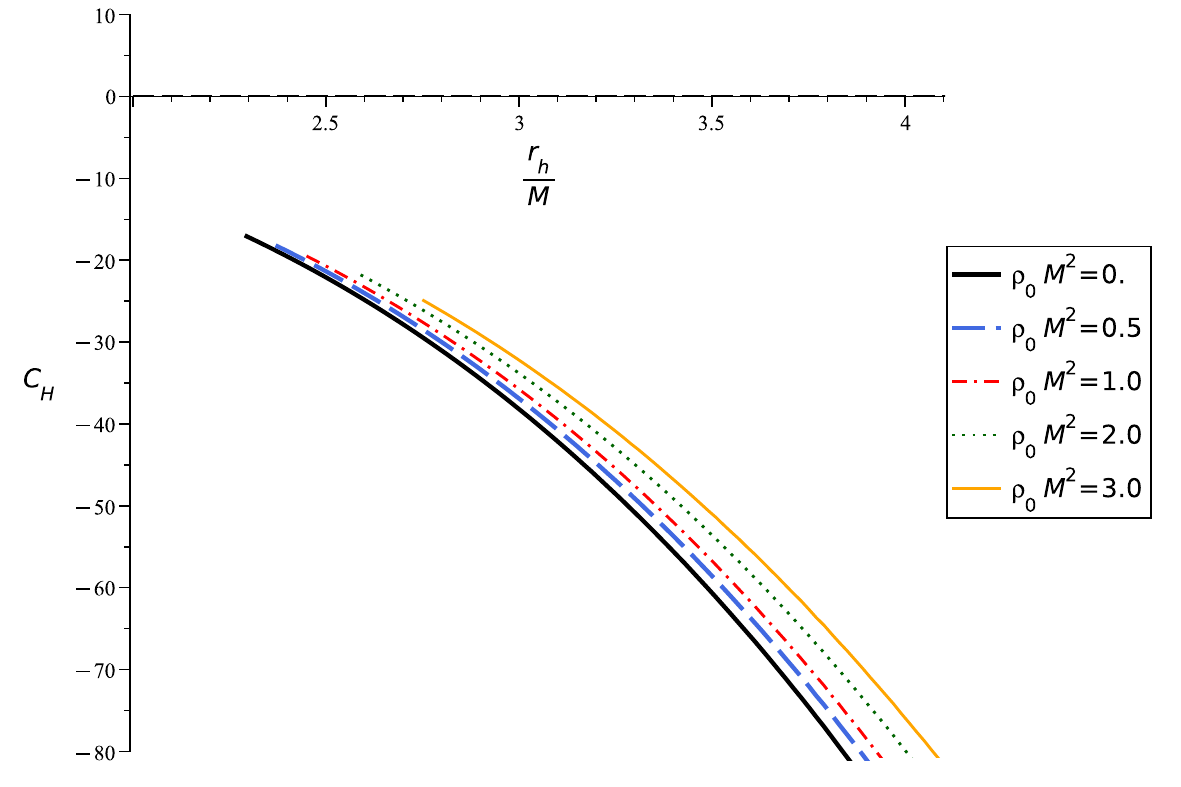}\\
    (ii) Varying $\rho_0$: $\alpha=0.1,\;r_0/M=0.2$
    \caption{Heat capacity $C_H$ vs $r_h/M$. Panel~(i): varying $\alpha$ at $\rho_0 M^2=0.5$. Panel~(ii): varying $\rho_0$ at $\alpha=0.1$. The heat capacity is negative throughout for all parameter choices, confirming that the Plummer-\CoS\ \BH\ is locally thermodynamically unstable. The dashed horizontal line marks $C_H=0$.}
    \label{fig:CH}
\end{figure}

A key finding of this work is that $C_{H}$ remains strictly negative for all values of $\rho_{0}$, $r_{0}$, and $\alpha<1$ studied (see Table~\ref{tab:thermo_data} and Fig.~\ref{fig:CH}). The Plummer-\CoS\ \BH\ is therefore locally thermodynamically unstable, just as the Schwarzschild \BH\ is. The absence of a sign change in $C_{H}$ can be traced back to the single-horizon, non-degenerate character of the solution established in Sec.~\ref{subsec:horizon} without a second horizon or an extremal limit, the denominator in~(\ref{eq:CH_def}) never vanishes, and there is no Davies-type phase transition~\cite{sec6is03}. This is in marked contrast to charged \BHs\ (Reissner-Nordstr\"{o}m) or AdS \BHs, where the heat capacity changes sign at a critical radius~\cite{sec6is02,sec6is03}.

Quantitatively, $|C_{H}|$ grows with $r_{h}$ roughly as $r_{h}^{2}$: at $r_{h}\approx 2M$ we find $C_{H}\approx -14$, whereas at $r_{h}=7M$ it reaches $C_{H}\approx -535$.

\subsection{Gibbs free energy and global stability}
\label{subsec:Gibbs}

The Gibbs free energy $G=M-T_{H}S$ characterizes the global thermodynamic preference of the \BH\ state relative to thermal radiation at the same temperature. Using Eqs.~(\ref{eq:mass}),~(\ref{eq:TH_explicit}), and~(\ref{eq:entropy}), one obtains
\begin{equation}
G=\frac{r_{h}}{4}\left[
\mathcal{C}\left(3+\mathcal{D}\right)-\alpha\right].
\label{eq:Gibbs}
\end{equation}
For the Schwarzschild \BH\ ($\rho_{0}=\alpha=0$) this reduces to $G=r_{h}/4=M/2>0$, which means the \BH\ is globally less stable than hot flat space-the well-known result underlying the Hawking-Page (HP) argument~\cite{sec6is04}.

\begin{table}[ht!]
\centering
\setlength{\tabcolsep}{6pt}
\renewcommand{\arraystretch}{1.3}
\begin{tabular}{|c|c|c|c|c|c|}
\hline
\rowcolor{orange!50}
\textbf{$\rho_0 M^2$} & \textbf{$\alpha$} & \textbf{$r_h/M$} & \textbf{$T_H M$} & \textbf{$C_H$} & \textbf{$G/M$} \\
\hline
0.0 & 0.00 & 2.050 & 0.03787 & $-13.53$ & 0.525 \\
0.0 & 0.00 & 3.000 & 0.01768 & $-42.42$ & 1.000 \\
0.0 & 0.00 & 7.000 & 0.00325 & $-539.3$ & 3.000 \\
\hline
0.5 & 0.00 & 2.123 & 0.03650 & $-14.54$ & 0.508 \\
0.5 & 0.00 & 3.073 & 0.01745 & $-44.06$ & 0.981 \\
\hline
0.0 & 0.10 & 2.272 & 0.03083 & $-16.58$ & 0.523 \\
0.0 & 0.10 & 3.222 & 0.01533 & $-47.30$ & 0.950 \\
\hline
0.5 & 0.10 & 2.354 & 0.02971 & $-17.83$ & 0.505 \\
0.5 & 0.10 & 7.304 & 0.00310 & $-530.8$ & 2.729 \\
\hline
0.5 & 0.30 & 3.014 & 0.01814 & $-29.07$ & 0.500 \\
0.5 & 0.30 & 4.964 & 0.00670 & $-129.8$ & 1.181 \\
\hline
0.0 & 0.50 & 4.050 & 0.00970 & $-52.16$ & 0.513 \\
0.0 & 0.50 & 9.000 & 0.00197 & $-573.8$ & 1.750 \\
\hline
\end{tabular}
\caption{Selected thermodynamic quantities for the Plummer-\CoS\ \BH\ at representative values of $\rho_0$, $\alpha$, and $r_h$, with $r_0/M=0.2$. The heat capacity $C_H$ is negative throughout, indicating thermodynamic instability for all parameter configurations.}
\label{tab:thermo_data}
\end{table}

\begin{figure}[ht!]
    \centering
    \includegraphics[width=0.65\linewidth]{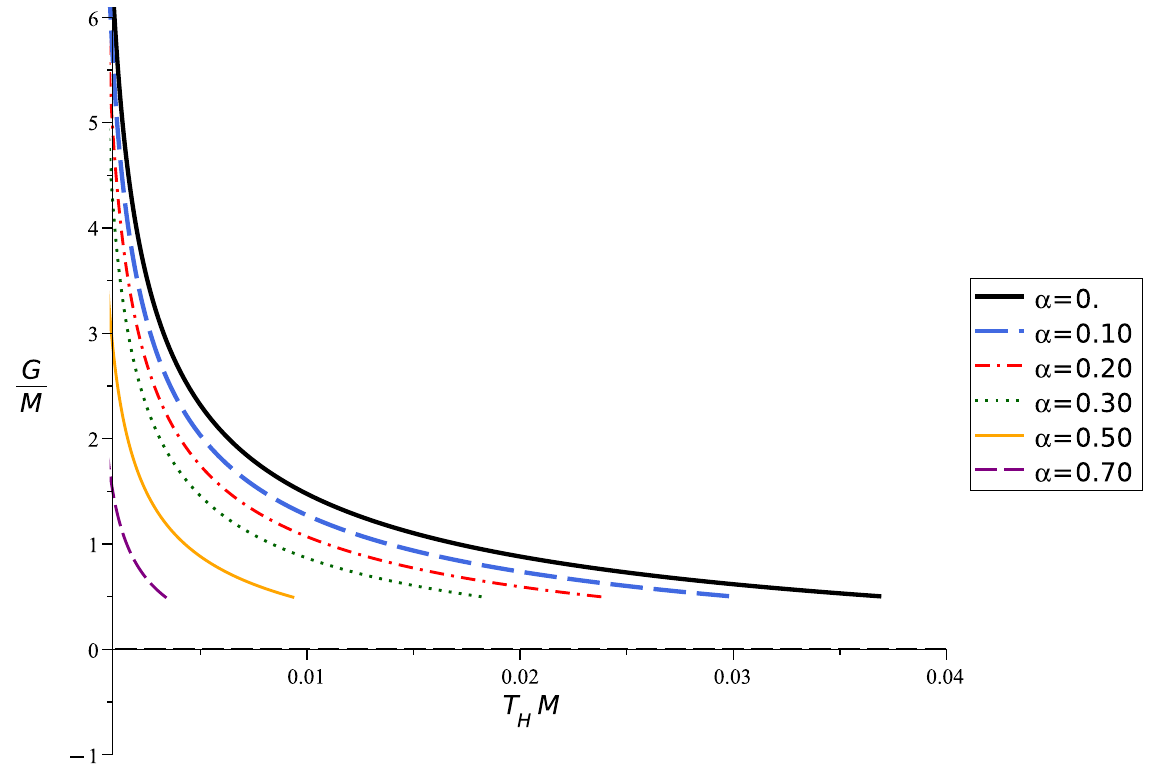}\\
    (i) Varying $\alpha$: $\rho_0 M^2=0.5,\;r_0/M=0.2$\\[3ex]
    \includegraphics[width=0.65\linewidth]{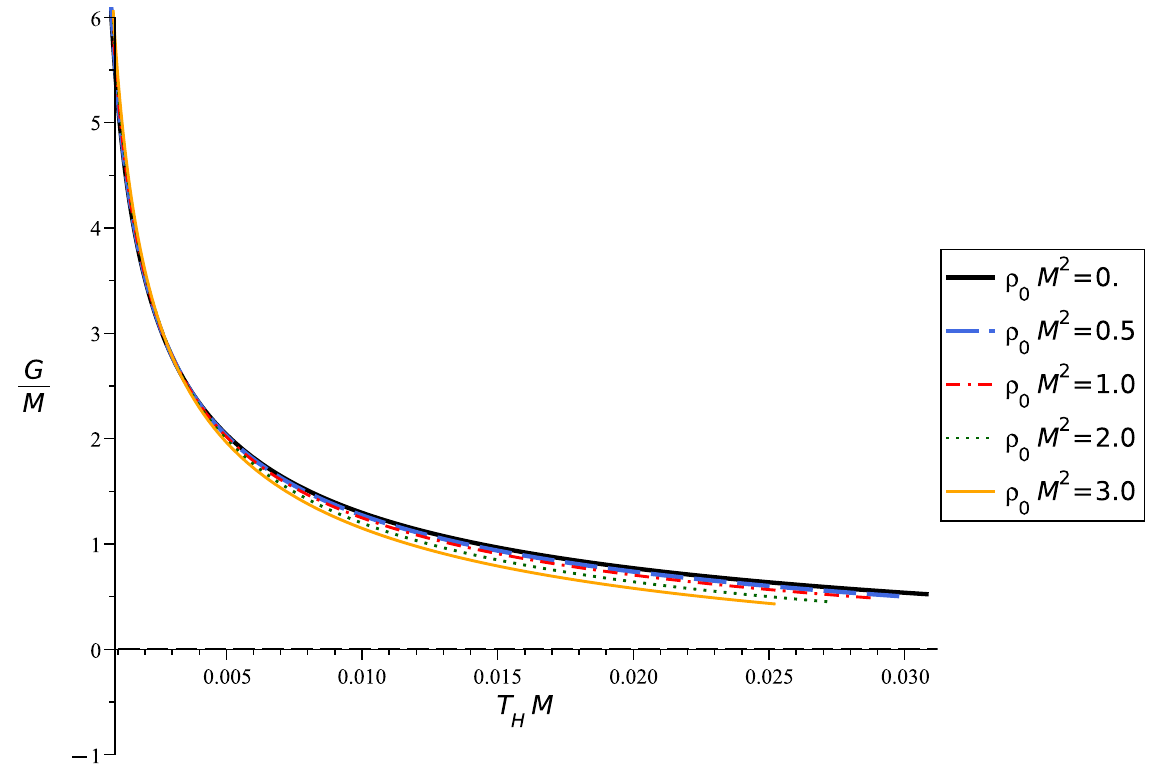}\\
    (ii) Varying $\rho_0$: $\alpha=0.1,\;r_0/M=0.2$
    \caption{Gibbs free energy $G/M$ vs Hawking temperature $T_H M$. Panel~(i): varying $\alpha$ at $\rho_0 M^2=0.5$. Panel~(ii): varying $\rho_0$ at $\alpha=0.1$. All curves lie above $G=0$ (dashed line), indicating the absence of an HP-like phase transition. Parameters: $r_0/M=0.2$.}
    \label{fig:Gibbs}
\end{figure}

Figure~\ref{fig:Gibbs} displays $G/M$ as a function of $T_{H}M$ for the two parameter scans. In panel~(i), $\alpha$ is varied at fixed $\rho_{0}M^{2}=0.5$: larger $\alpha$ shifts the Gibbs curves downward and to the left, producing a colder \BH\ at a given $G/M$ and moving the endpoint toward the origin. In panel~(ii), $\rho_{0}$ is varied at fixed $\alpha=0.1$: the shift is similar in direction but weaker in magnitude. In all cases $G>0$ throughout the accessible temperature range, and no swallow-tail structure appears. The absence of a negative-$G$ branch confirms that no HP-like first-order transition occurs in this geometry. This is physically expected: the HP transition requires an effective confining mechanism (such as an AdS box) to stabilize large \BHs; in the present asymptotically conical-deficit spacetime no such mechanism is available. In the full $(\rho_{0},\alpha)$ parameter space the Gibbs free energy is always positive, reinforcing the conclusion of thermodynamic instability drawn from the heat capacity analysis.

\section{Conclusion}\label{sec:7}

We have constructed a static, spherically symmetric \BH\ solution embedded in a cored Plummer \DM\ halo and threaded by a Letelier \CoS, extending the Plummer-Schwarzschild metric of Ref.~\cite{Senjaya2026} by incorporating the string-cloud tension parameter $\alpha$ into the lapse function. The resulting metric function $A(r) = \exp[-4\pi\rho_{0}r_{0}^{3}\tan^{-1}(r/r_{0})/r] - r_{s}/r - \alpha$ was analyzed across six interconnected physical domains, and the principal findings are summarized below.

The spacetime possesses a single, non-degenerate \EH\ for all $\alpha < 1$, with no inner horizon or extremal limit. The \EH\ radius grows with both the halo central density $\rho_{0}$ and the \CoS\ tension $\alpha$, from the Schwarzschild value $r_{h}=2M$ at $\rho_{0}=\alpha=0$ to $r_{h}\approx 41.6\,M$ at $(\rho_{0}M^{2},\alpha)=(0.5,\,0.95)$, diverging as $\alpha\to 1^{-}$ according to $r_{h}\approx 2M/(1-\alpha)$. For $\alpha\ge 1$ no horizon exists and the metric describes a naked singularity.

The \PS\ radius, shadow radius, and \ISCO\ all increase monotonically with $\rho_{0}$ and $\alpha$, following the same hierarchy: the \CoS\ tension produces the dominant shift, while the Plummer halo contributes a secondary correction. At $(\rho_{0}M^{2},\alpha)=(0.5,\,0.3)$, the shadow radius reaches $R_{\rm sh}\approx 8.83\,M$ (versus $5.20\,M$ for Schwarzschild) and the \ISCO\ expands to $r_{\rm ISCO}\approx 12.27\,M$ (versus $6M$), more than doubling the Schwarzschild values.

The weak deflection angle, computed via the \GBT, receives two distinct types of corrections: the \CoS\ rescales the entire Einstein deflection by a factor $(1-\alpha)^{-1}$, while the Plummer halo adds a genuine mass-like contribution proportional to $\rho_{0}r_{0}^{3}$. The different functional dependences on $\alpha$-namely $\sqrt{1-\alpha}$ for the shadow and $(1-\alpha)^{-1}$ for the deflection angle-offer a potential route to disentangling the two effects from independent measurements.

The scalar perturbation potential $V_{s}(r)$ is lowered and broadened by both $\rho_{0}$ and $\alpha$, leading to enhanced \GFs\ (higher transmission through the barrier) relative to the Schwarzschild baseline. The Boonserm-Visser bounds confirm a $\sim 22\%$ increase in the $\ell=0$ \GF\ at $\omega M = 0.2$ for $(\rho_{0}M^{2},\alpha)=(0.5,\,0.1)$. Despite this enhanced transparency, the net Hawking emission rate is suppressed because the reduced Hawking temperature $T_{H}$ weakens the thermal factor more than the \GF\ enhancement can compensate.

The \QNM\ spectrum, obtained via the first-order \WKB\ method, shows that both the oscillation frequency $\omega_{R}$ and the damping rate $|\omega_{I}|$ decrease with increasing $\rho_{0}$ and $\alpha$. For $\ell=2$, $n=0$, the oscillation frequency drops by $45\%$ and the damping rate by $53\%$ as $\alpha$ increases from $0$ to $0.3$ (at $\rho_{0}M^{2}=0.5$), while the quality factor $\mathcal{Q}=|\omega_{R}/(2\omega_{I})|$ rises from $2.63$ to $3.09$, indicating a more monochromatic ringdown signal.

The thermodynamic analysis reveals that the Hawking temperature is positive and monotonically decreasing in $r_{h}$ for all parameter configurations. The heat capacity $C_{H}$ remains strictly negative throughout the accessible parameter space, confirming that the Plummer-\CoS\ \BH\ is locally thermodynamically unstable-analogous to the Schwarzschild case and in contrast to charged or AdS \BHs\ that exhibit Davies-type phase transitions. The Gibbs free energy $G$ is positive for all temperatures, ruling out any Hawking-Page-like first-order transition, which is consistent with the absence of an effective confining mechanism in the asymptotically conical-deficit geometry.

A unifying theme across all six analyses is the hierarchical parameter dependence: the \CoS\ tension $\alpha$ governs the leading-order modifications to every observable, while the Plummer halo density $\rho_{0}$ provides a subdominant, additive correction. This hierarchy originates from the structural roles of the two parameters in the metric function--$\alpha$ enters as a constant shift that alters the asymptotic value $A(\infty)=1-\alpha$ and globally rescales the geometry, whereas the \DM\ contribution is exponentially suppressed and localized near the core radius $r_{0}$.

\scriptsize

\section*{Acknowledgments}

F.A. gratefully acknowledges the Inter University Centre for
Astronomy and Astrophysics (IUCAA), Pune, India, for the
opportunity to serve as a visiting associate.
\.{I}.~S. expresses his thanks to T\"{U}B\.{I}TAK, ANKOS,
and SCOAP3 for their financial support.
He further recognizes the backing of COST Actions CA22113,
CA21106, CA23130, CA21136, and CA23115, which have played a
important role in strengthening networking activities.

\section*{Data Availability Statement}

In this study, no new data was generated or analyzed.

\scriptsize

\bibliographystyle{apsrev4-2}

\bibliography{ref}

\end{document}